\documentclass[epj]{svjour}
\usepackage[dvips]{graphics}
\usepackage{dcolumn}
\usepackage{amsmath}
\usepackage{latexsym}
\usepackage{longtable}
\usepackage{amssymb}
\begin{document}
\title{Signals on the power spectra of a cosmology modeled with Chebyshev polynomials}
\author{Freddy Cueva Solano\inst{1}
\thanks{\emph{e-mail:freddy@ifm.umich.mx}} 
}                
\institute{Instituto de F\'{\i}sica y Matem\'aticas, Universidad Michoacana de San Nicol\'as de Hidalgo\\
Edificio C-3, Ciudad Universitaria, CP. 58040, Morelia, Michoac\'an, M\'exico}
\date{\today}
\abstract{
We present an interacting model with a phenomenological interaction, $\bar{Q}$, between a cold dark matter (DM) fluid and a dark 
energy (DE) fluid, which takes a time-varying equation of state (EoS) parameter, $\mathrm{\omega_{DE}}$. Here, both $\bar{Q}$ and 
$\mathrm{\omega_{DE}}$ are modeled in terms of the Chebyschev polynomials. In a Newtonian gauge and on sub-horizon scales, a set of perturbed 
equations is obtained when the momentum transfer potential becomes null in the DM rest-frame. This leads to different cases of the interacting model. 
Then, via a Markov-Chain Monte Carlo (MCMC) method, we constrain such cases by using a combined analysis of geometric and dynamical data. 
Our results show that in such cases the evolution curves of the structure growth of the matter deviate strongly from the standard model. 
In addition, we also found that the matter power spectrum is sensitive to $\bar{Q}$. In this way, the coupling modifies the matter scale and 
generates a slight variation of the turnover point to smaller scales. Likewise, the amplitude of the CMB temperature power spectrum is sensitive 
the values of $\bar{Q}$ and $\mathrm{\omega_{DE}}$ at low and high multipoles $\it l$, respectively. Here, $\bar{Q}$ can cross twice the line 
$\bar{Q}=0$ during its background evolution.
} 
\maketitle 
\section{Introduction} 
A number of observations 
\cite{Conley2011,Jonsson2010,Betoule2014,Jackson1972,Kaiser1987,Mehrabi2015,Alcock1979,Seo2008,Battye2015,Samushia2014,Hudson2013,Beutler2012,Feix2015,Percival2004,Song2009,Tegmark2006,Guzzo2008,Samushia2012,Blake2011,Tojeiro2012,Reid2012,delaTorre2013,Planck2015,Hinshaw2013,Beutler2011,Ross2015,Percival2010,Kazin2010,Padmanabhan2012,Chuang2013a,Chuang2013b,Anderson2014a,Kazin2014,Debulac2015,FontRibera2014,Eisenstein1998,Eisenstein2005,Hemantha2014,Bond-Tegmark1997,Hu-Sugiyama1996,Neveu2016,Sharov2015,Zhang2014,Simon2005,Moresco2012,Gastanaga2009,Oka2014,Blake2012,Stern2010,Moresco2015,Busca2013} have indicated that the present universe is undergoing a phase of accelerated expansion, and 
driven probably by a new form of energy with negative EoS parameter, commonly so-called DE \cite{DES2006}. This energy has been interpreted in 
various forms and widely studied in \cite{OptionsDE}. 
However, within General Relativity (GR) the DE models can suffer the coincidence problem, namely why the DM and DE energy densities are of the same 
order today. This latter problem could be solved or even alleviated, by assuming the existence of a non-gravitational $\bar{Q}$ within the dark 
sector, which gives rise to a continuous energy exchange from DE to DM or vice-versa. Currently, there are n't neither physical arguments nor 
recent observations to exclude $\bar{Q}$ \cite{Interacting,Pavons,Wangs,Cueva-Nucamendi2012}. Moreover, due to the absence of a fundamental theory 
to construct $\bar{Q}$, different ansatzes have been widely discussed in \cite{Interacting,Pavons,Wangs,Cueva-Nucamendi2012,valiviita2008,Clemson2012}. 
So, It has been shown in some coupled DE scenarios that $\bar{Q}$ can affect the background evolution of the DM density perturbations and the expansion 
history of the universe \cite{Mehrabi2015,valiviita2008,Clemson2012,Alcaniz2013,Yang2014}. Thus, $\bar{Q}$ and $\mathrm{\omega_{DE}}$ could very possibly introduce new features on 
the evolution curves of the structure growth of the matter, on the linear matter power spectrum and on the amplitude of the CMB temperature power 
spectrum at low and high multipoles, respectively \cite{Mehrabi2015,Clemson2012,Alcaniz2013,Yang2014,Mota2017}. \\
On the other one, within dark sector we can propose new ansatzes for both $\bar{Q}$ and $\mathrm{\omega_{DE}}$, which can be expanded in terms 
of the Chebyshev polynomials $T_{n}$, defined in the interval $[-1,1]$ and with a divergence-free $\mathrm{\omega_{DE}}$ at z$\rightarrow -1$ \cite{Chevallier-Linder,Li-Ma}. 
However, that polynomial base was particularly chosen due to its rapid convergence and better stability than others, by giving minimal 
errors \cite{Simon2005,Martinez2008}. Besides, $\bar{Q}$ could also be proportional to the DM energy density $\mathrm{{\bar{\rho}}_{DM}}$ 
and to the Hubble parameter $\bar{H}$. This new model will guarantee an accelerated scaling attractor and connect to a standard 
evolution of the matter. Here, $\bar{Q}$ will be res-tricted from the criteria exhibit in \cite{Campo-Herrera2015}.\\
The focus of this paper is to investigate the effects of $\bar{Q}$ and $\mathrm{\omega_{DE}}$ on the curves of structure growth, on the ma-tter power 
spectrum and on the CMB temperature power spectrum including the search for a new way to alleviate the coincidence problem.\\
On the other hand, an interacting DE model is discussed, on which we have performed a global fitting, by using an analysis combined of Joint Light Curve 
Analysis (JLA) type Ia Supernovae (SNe Ia) data \cite{Conley2011,Jonsson2010,Betoule2014}, including the growth rate of structure formation 
obtained from redshift space distortion (RSD) data 
\cite{Jackson1972,Kaiser1987,Mehrabi2015,Alcock1979,Seo2008,Battye2015,Samushia2014,Hudson2013,Beutler2012,Feix2015,Percival2004,Song2009,Tegmark2006,Guzzo2008,Samushia2012,Blake2011,Tojeiro2012,Reid2012,delaTorre2013,Planck2015}, 
together with Baryon Acoustic Oscillation (BAO) data 
\cite{Hinshaw2013,Beutler2011,Ross2015,Percival2010,Kazin2010,Padmanabhan2012,Chuang2013a,Chuang2013b,Anderson2014a,Kazin2014,Debulac2015,FontRibera2014,Eisenstein1998,Eisenstein2005,Hemantha2014}, 
as well as the observations of anisotropies in the power spectrum of the Cosmic Microwave Background (CMB) data 
\cite{Planck2015,Bond-Tegmark1997,Hu-Sugiyama1996,Neveu2016} and the Hubble parameter (H) data obtained from galaxy 
surveys \cite{Sharov2015,Zhang2014,Simon2005,Moresco2012,Gastanaga2009,Oka2014,Blake2012,Stern2010,Moresco2015,Busca2013} to constrain the 
parameter space of such model and break the degeneracy of their parameters, putting tighter constraints on them.\\
Finally, we organize this paper as follows: We describe the background equations of the interacting DE model in Sec. II, the perturbed equations, 
the modified growth factor, the linear matter and CMB temperature power spectra in Sec. III. The constraint method and observational data are presented in 
Sec. IV. We discuss our results in Sec. V and show our conclusions in Sec. VI.
\section{Interacting dark energy (IDE) model}\label{Background}
We assume a spatially flat Friedmann-Robertson-Walker (FRW) universe, composed with four perfect fluids-like, radiation (subscript r), baryonic matter (subscript b), 
DM and DE, respectively. Moreover, we postulate the existence of a non-gravitational coupling in the background between DM and DE (so-called dark sector) and two 
decoupled sectors related to the b and r components, respectively. We also consider that these fluids have EoS parameters ${\mathrm{P}}_{A}=\mathrm{\omega}_{A}\mathrm{\bar{\rho}}_{A}$, 
$A=b, r, DM, DE$, where $\mathrm{P}_{A}$ and $\mathrm{{\bar{\rho}}}_{A}$ are the corresponding pressures and the energy densities. Here, we choose $\mathrm{\omega_{DM}=\omega_{b}=0}$, 
$\mathrm{\omega_{r}=1/3}$ and $\mathrm{\omega_{DE}}$ is a time-varying function. Therefore, the balance equations of our fluids are respectively, 
\begin{eqnarray}
\label{EB}\frac{\mathrm{d}\bar{\rho}_{b}}{\mathrm{d}z}-3{\bar{H}}{\bar{\rho}}_{b}&=&0,\\
\label{Er}\frac{\mathrm{d}\bar{\rho}_{r}}{\mathrm{d}z}-4{\bar{H}}{\bar{\rho}}_{r}&=&0,\\
\label{EDM}\frac{\mathrm{d}\bar{\rho}_{DM}}{\mathrm{d}z}-\frac{3{\bar{\rho}}_{DM}}{(1+z)}&=&-\frac{\bar{Q}}{\bar{H}(1+z)},\\
\label{EDE}\frac{\mathrm{d}\bar{\rho}_{DE}}{\mathrm{d}z}-\frac{3(1+\omega_{DE}){\bar{\rho}}_{DE}}{(1+z)}&=&+\frac{\bar{Q}}{\bar{H}(1+z)},
\end{eqnarray}
where the differentiation has been done with respect to the redshift, $\mathrm{z}$, $\bar{H}$ denotes the Hubble expansion rate and the quantity $\bar{Q}$ expresses the 
interaction between the dark sectors. For simplicity, it is convenient to define the fractional energy densities 
$\mathrm{{\bar{\Omega}}_{i}}\equiv\frac{\mathrm{\bar{\rho}_{A}}}{\rho_{c}}$ and ${\mathrm{\Omega}}_{A,0}\equiv\frac{{\rho}_{i,0}}{\rho_{c,0}}$, where the critical density 
$\rho_{c}\equiv 3{\bar{H}}^2/8\pi G$ and the critical density today $\rho_{c,0}\equiv 3\mathrm{H_{0}}^{2}/8\pi G$ being $\mathrm{H_{0}}=100h\,Kms^{-1}Mpc^{-1}$ the current value of $\bar{H}$. 
Likewise, we have taken the relation $\mathrm{{\bar{\Omega}}_{b,0}+{\bar{\Omega}}_{r,0}+{\bar{\Omega}}_{DM,0}+{\bar{\Omega}}_{DE,0}=1}$. Here, the subscript ``0'' indicates the 
present day value of the quantity.\\ 
In this work, we consider the spatially flat FRW metric with line element 
\begin{equation}\label{metricbackground} 
{\rm ds}^{2}=-{\rm d}{t}^{2}+{\mathrm{\bf a}}^{2}(t)\delta_{ij}d{x}^{i}d{x}^{j},
\end{equation}
where $t$ represents the cosmic time and ``$\mathrm{\bf a}$'' represents the scale factor of the metric and it is defined in terms of the 
redshift $\mathrm{z}$ as $\mathrm{\bf a}=(1+\mathrm{z})^{-1}$.\\
Here, we analyze the ratio between the energy densities of DM and DE, defined as $\mathrm{R}\equiv \mathrm{{\bar{\rho}}_{DM}}/\mathrm{{\bar{\rho}}_{DE}}$. 
From Eqs. (\ref{EDM}) and (\ref{EDE}), we obtain \cite{Campo-Herrera2015,Ratios}
\begin{equation}\label{ratio}
\frac{\mathrm{d}R}{\mathrm{d}z}=\frac{-R}{(1+z)}\left(3\omega_{DE}+\frac{(1+R)\bar{Q}}{\bar{H}\rho_{DM}}\right).
\end{equation}
This Eq. leads to 
\begin{equation}\label{FormQ}
\bar{Q}=-\left(3\omega_{DE}+\frac{\mathrm{d}R}{\mathrm{d}z}\frac{(1+z)}{R}\right)\frac{\bar{H}{\bar{\rho}}_{DM}}{1+R}.
\end{equation}
Due to the fact that the origin and nature of the dark fluids are unknown, it is not possible to derive $\bar{Q}$ from fundamental principles. 
However, we have the freedom of choosing any possible form of $\bar{Q}$ that satisfies Eqs. (\ref{EDM}) and (\ref{EDE}) simultaneously. Hence, we 
propose a phenomenological description for $\bar{Q}$ as a linear combination of $\mathrm{{\bar{\rho}}_{DM}}$, ${\bar{H}}$ and a time-varying 
function ${\bar{\rm I}}_{\rm Q}$,
\begin{equation}\label{Interaction}
{\bar{Q}}\equiv \mathrm{\bar{H}}\mathrm{\bar{\rho}_{DM}}\bar{{\rm I}}_{\rm Q},\qquad {\bar{\rm I}}_{\rm Q}\equiv \sum_{n=0}^{}{{\lambda}}_{n}T_{n},
\end{equation}
where ${\bar{\rm I}}_{\rm Q}$ is defined in terms of Chebyshev polynomials and $\mathrm{{\lambda}_{n}}$ are constant and small $|\mathrm{{\lambda}_{n}}|\ll1$ 
dimensionless parameters. This polynomial base was chosen because it converges rapidly, is more stable than others and behaves well in any polynomial 
expansion, giving minimal errors \cite{Cueva-Nucamendi2012}. The first three Chebyshev polynomials are
\begin{equation}\label{Chebyshev1} 
T_{0}(z)=1\;,\hspace{0.3cm}T_{1}(z)=z\;,\hspace{0.3cm} T_{2}(z)=(2z^{2}-1).
\end{equation}
From Eqs. (\ref{Interaction}) and (\ref{Chebyshev1}) an asymptotic value for ${\bar{\rm I}}_{\rm Q}$ can be found:
${\bar{\rm I}}_{\rm Q}\rightarrow \infty$ for $\mathrm{z}\rightarrow \infty$, ${\bar{\rm I}}_{\rm Q}=\mathrm{{\lambda}_{0}-{\lambda}_{2}}$ for $\mathrm{z}=0$ and 
${\bar{\rm I}}_{\rm Q}\approx \mathrm{{\lambda}_{0}-{\lambda}_{1}+{\lambda}_{2}}$ for $\mathrm{z}\rightarrow -1$.\\
Similarly, we will focus on an interacting model with a specific ansatz for the EoS parameter, given as
\begin{equation}\label{wde}
\mathrm{\omega_{DE}}\equiv \omega_{2}+2\sum^{2}_{m=0}\frac{\omega_{m}T_{m}}{2+{\mathrm{z}}^{2}}. 
\end{equation}
Within this ansatz a finite value for $\omega$ is obtained from the past to the future; namely, the following asymptotic values are found: 
$\mathrm{\omega_{DE}}=5\omega_{2}$ for $\mathrm{z}\rightarrow \infty$, $\mathrm{\omega_{DE}}\approx \omega_{0}$ for $\mathrm{z}=0$ and 
$\mathrm{\omega_{DE}}\approx(5/3)\omega_{2}+(2/3)[\omega_{0}-\omega_{1}]$ for $\mathrm{z}\rightarrow -1$. Therefore, a possible physical description 
should be studied to explore its properties.\\
In order to guarantee that $\bar{Q}$ may be physically acceptable in the dark sectors \cite{Campo-Herrera2015}, we equal the right-hand sides of Eqs. (\ref{FormQ}) and 
(\ref{Interaction}), which becomes
\begin{equation}\label{criterion}
\frac{\mathrm{d R}}{\mathrm{d}z}=\frac{-\mathrm{R}}{(1+\mathrm{z})}\biggl({\bar{\rm I}}_{\rm Q}(1+\mathrm{R})+3\mathrm{{\omega}_{DE}}\biggr). 
\end{equation}
Now, to solve or alleviate of coincidence problem, we require that $\mathrm{R}$ tends to a fixed value at late times. This leads to the condition
${\mathrm{d R}}/{\mathrm{d z}}=0$, which therefore implies two stationary solutions $\mathrm{R_{+}}=\mathrm{R(z\rightarrow\infty)}=
-(1+{3\mathrm{\omega_{DE}}}/{{\bar{\rm I}}_{\rm Q}})$ and $\mathrm{R_{-}}=\mathrm{R(z\rightarrow -1)}=0$,
The first solution occurs in the past and the second one happens in the future.\\
By inserting Eqs. (\ref{Interaction}) and (\ref{wde}) into Eq. (\ref{criterion}), we find that $\mathrm{R}$ has no analytical solution, in any case, it is to be solved 
numerically. Likewise, there are an analytical solution for just $\mathrm{{\bar{\rho}}_{b}}$, $\mathrm{{\bar{\rho}}_{r}}$ and $\mathrm{{\bar{\rho}}_{DM}}$, respectively, 
but $\mathrm{{\bar{\rho}}_{DE}}$ will be obtained from $\mathrm{R}$, as $\mathrm{{\bar{\rho}}_{DE}}=\mathrm{{\bar{\rho}}_{DM}}/\mathrm{R}$.\\ 
Therefore, the first Friedmann equation is given by
\begin{equation}\label{hubble}
\smallskip
E^{2}=\frac{{\bar{H}}^{2}}{\mathrm{{H}^{2}_{0}}}=\mathrm{\Omega_{b,0}}{(1+\mathrm{z})}^{3}+\mathrm{\Omega_{r,0}}{(1+\mathrm{z})}^{4}\\
+\mathrm{\Omega^{\star}_{DM}}(\mathrm{z})(1+{\mathrm{R}}^{-1}),
\end{equation}
where have considered that
\begin{eqnarray}
\smallskip
\mathrm{\Omega^{\star}_{DM}}(\mathrm{z})=(1+\mathrm{z})^{3}{\mathrm{\Omega_{DM,0}}}{\rm exp}\biggl[{\frac{-z_{max}}{2}\sum_{n=0}^{2}\lambda_{n}I_{n}(\mathrm{z})}\biggr],\qquad\qquad\qquad\nonumber\\
\int_{0}^{\mathrm{z}}\frac{T_{n}(\tilde{x})}{(1+\tilde{x})}d\tilde{x}\,\approx\,\frac{z_{max}}{2}\int_{-1}^{x}\frac{T_{n}(\tilde{x})}{(a_{1}+a_{2}\tilde{x})}d\tilde{x}\equiv\frac{z_{max}}{2}I_{n}(\mathrm{z}),\qquad\qquad\nonumber\\
x\,\equiv\, \frac{2\mathrm{z}}{z_{max}}-1,\quad a_{1}\equiv1+\frac{z_{max}}{2}\,,\quad a_{2}\equiv\frac{z_{max}}{2},\qquad\qquad\qquad\nonumber\\
I_{0}(\mathrm{z})=\frac{2}{z_{max}}\ln(1+\mathrm{z}),\qquad\qquad\qquad\qquad\qquad\qquad\qquad\qquad\qquad\nonumber\\
I_{1}(\mathrm{z})=\,\frac{2}{z_{max}}\biggl(\frac{2\mathrm{z}}{z_{max}}-\frac{(2+z_{max})}{z_{max}}\ln(1+\mathrm{z})\biggr),\qquad\qquad\qquad\qquad\nonumber\\
I_{2}(\mathrm{z})\,=\,\frac{2}{z_{max}}\biggl[\frac{4 \mathrm{z}}{z_{max}}\biggl(\frac{\mathrm{z}}{z_{max}}-\frac{2}{z_{max}}-2\biggr)+\qquad\qquad\qquad\qquad\quad\nonumber\\
\biggl(1+\frac{6.8284}{z_{max}}\biggr)\biggl(1+\frac{1.1716}{z_{max}}\biggr)\ln(1+\mathrm{z})\biggr],\nonumber\hspace{3cm}
\end{eqnarray}
where $z_{max}$ is the maximum value of $\mathrm{z}$ such that $\tilde{x}\in [-1, 1]$ and $\vert{T_{n}}(\tilde{x})\vert\leq1$ and $n\in[0,2]$ 
\cite{Cueva-Nucamendi2012}.\\ 
If $\bar{Q}(z)=0$ and $\mathrm{\omega_{DE}}=-1$ in Eq. (\ref{hubble}) the standard $\Lambda$CDM model is recovered. Similarly, when $\bar{Q}(z)=0$ and 
$\mathrm{\omega_{DE}}$ are nonzero, the $\omega$DE model is obtained. These non-interacting models have an analytical solution for $\mathrm{R}$. 
\section{IDE in the perturbed universe}\label{structure} 
\subsection{Perturbed equations}\label{structure}
In the Newtonian gauge, the perturbed FRW metric becomes \cite{valiviita2008,Clemson2012}
\begin{equation}\label{metricFRW} 
{\rm ds}^{2}=\biggl[-\left(1+2\phi\right){\rm d}{t}^{2}+{\mathrm{\bf a}}^{2}(t)\left(1-2\psi\right)\delta_{ij}d{x}^{i}d{x}^{j}\biggr],
\end{equation}  
where $\phi$ and $\psi$ are gravitational potentials, and the four-velocity of fluid $A$ ($A=$ DM, DE, b, r) is 
\begin{equation}\label{fourvelocity}
U^{\mu}_{A}={\mathrm{\bf a}}^{-1}(1-\phi,\partial^{i}v_{A})\,,\quad U_{\mu}^{A}={\mathrm{\bf a}}(-1-\phi,\partial_{i}v_{A}),
\end{equation}
where $v_{A}$ is the peculiar velocity potential, and $\theta_{A}$ is the velocity perturbation defined as $\theta_{A}=-k^{2}v_{A}$.\\
The energy-momentum conservation equation of $A$ fluid in interaction is given by \cite{valiviita2008,Clemson2012}
\begin{equation}\label{Energy_momentum_transferA} 
\nabla_{\nu}{T_{A}}^{\mu\nu}=Q_{A}^{\mu},\,\,\, Q_{DM}^{\mu}=-Q_{DE}^{\mu}\neq 0=Q_{b}^{\mu}=Q_{r}^{\mu},
\end{equation}
where ${T_{A}}^{\mu\nu}$ is the $A$-fluid energy momentum tensor.\\
In general, $Q_{A}^{\mu}$ can be split relative to the total four-velocity $U^{\mu}$ as \cite{valiviita2008,Clemson2012}
\begin{eqnarray} \label{decomposeQA}
Q_{A}^{\mu}&=&{Q}_{A}U^{\mu}_{A}+F^{\mu}_{A},\quad Q_{A}=\bar{Q}_{A}+\delta Q_{A},\quad U_{\mu}^{A}F^{\mu}_{A}=0,\nonumber\\
\bar{Q}_{DM}&=&-\bar{Q}_{DE}\neq 0=\bar{Q}_{b}=-\bar{Q}_{r},\nonumber\\
\delta{Q}_{DM}&=&-\delta{Q}_{DE}\neq 0=\delta{Q}_{b}=-\delta{Q}_{r},
\end{eqnarray}
where ${Q}_{A}$ and $F^{\mu}_{A}$ represent the energy and momentum transfer rate, respectively, relative to $U^{\mu}_{A}$. Likewise, to the first order 
$F^{\mu}_{A}={\mathrm{ \bf a}}^{-1}(0,\partial^{i}f_{A})$ where $f_{A}$ is a momentum transfer potential and $\bar{Q}_{A}$ represents the interaction term. 
From Eqs. (\ref{fourvelocity}) and (\ref{decomposeQA}), we find \cite{Clemson2012}
\begin{equation}\label{componentF}
Q^{A}_{0}=-{\mathrm{\bf a}}\left[\bar{Q}_{A}(1+\phi)+\delta{Q_{A}}\right],\quad 
Q^{A}_{i}={\mathrm{\bf a}}\partial_{k}\left[f_{A}+\bar{Q}_{A}v\right],  
\end{equation}
with $\bar{Q}_{DM}=-\bar{Q}_{DE}\neq 0$ and $\delta{Q}_{DM}=-\delta{Q}_{DE}$.\\
Here, we have considered that the $A$ fluid physical sound speed in the rest-frame defined by 
$c_{sA}^{2}\equiv\delta{{P}_{A}}/\delta{\rho}_{A}{\mid}_{rf}$ 
and the adiabatic sound speed is defined by 
$c_{{\bf \mathrm{a}}A}^{2}\equiv {\partial{P}_{A}}/{\partial{\rho}_{A}}=\mathrm{\omega_{A}}+
(\frac{\mathrm{d}{\omega}_{A}}{\mathrm{d}z}/\frac{\mathrm{d}{\bar{\rho}_{A}}}{\mathrm{d}z})\mathrm{\bar{\rho}_{A}}$
Then, for the adiabatic DM fluid, we take $c^{2}_{sDM}=c^{2}_{{\bf \mathrm{a}}DM}=\mathrm{\omega_{DM}}=0$. Instead, for the non-adiabatic DE fluid, 
$c^{2}_{\mathrm{aDE}}=\mathrm{\omega_{DE}}<0$ and the physical sound speed for DE is usually considered as $c^{2}_{sDE}=1$ to eliminate possible 
unphysical instabilities.\\
Immediately, we have established the simpler physical choice for the momentum transfer potential between the dark sectors, which happens 
when $f_{A}=0$  in the rest-frame of either DM or DE \cite{valiviita2008}. Consequently, this choice allows two different possibilities 
for $Q_{A}^{\mu}$ and $f_{A}$, which can be parallel to either the DM or the DE four velocity, respectively. In this work, we focus only on 
the case 
\cite{valiviita2008,Clemson2012}
\begin{eqnarray}
\label{CovariantQ1} 
{Q}_{DE}^{\mu}={\bar{Q}}_{DE}U^{\mu}_{DM}&=&-{Q}_{DM}^{\mu},\quad Q^{\mu}_{A}\parallel U^{\mu}_{DM},\qquad\nonumber\\
f_{DM}=\frac{{\bar{Q}}_{A}}{k^{2}}(\theta-\theta_{DM})&=&-f_{DE}\,,\,\quad Q^{\mu}_{A}\parallel U^{\mu}_{DM},
\end{eqnarray}
On the other hand, assuming that $\bar{Q}$ depends on the cosmic time through the global expansion rate, then a possible choice for 
$\delta{\bar{H}}$ can be $\delta{\bar{H}}=0$. Likewise, for convenience, we impose that ${{\delta{\rm I}}_{\rm Q}}\ll{\delta_{DM}}$, it leads to
\begin{equation}\label{parts1_Exchange_energy}
{\delta Q}_{DM}=-{\bar{H}}{\bar{\rm I}}_{\rm Q}\mathrm{{\bar{\rho}}_{DM}}{\delta}_{DM}\;.
\end{equation}
In a forthcoming article we will extend our study, by considering other relations between ${{\delta{\rm I}}_{\rm Q}}$, ${\delta_{DM}}$ and $\delta{H}$. 
It is beyond the scope of the present paper.\\ 
In this work, we are interested in studying the effects of $\mathrm{\omega_{DE}}$ and $\bar{Q}$ on the total matter power spectrum and on the CMB 
temperature power spectrum. For this reason, we consider only the adiabatic perturbations, assume that ${T_{A}}^{\mu\nu}$ is free of anisotropic 
stress, and the arguments above discussed, we find the evolution equations for the density contrast perturbation 
$\delta_{A}$ and the velocity perturbations $\theta_{A}$ in the IDE model from the general case presented in \cite{valiviita2008,Clemson2012} 
when $Q^{\mu}_{A}\parallel U^{\mu}_{DM}$,  
\begin{eqnarray}
\label{contrastDE}
\frac{\mathrm{d}{\delta}_{DE}}{\mathrm{d}z}&=&-(\mathrm{R}{\bar{\rm I}}_{\rm Q}-3+3\mathrm{\omega_{DE}})\frac{\delta_{DE}}{(1+\mathrm{z})}
+(1+\mathrm{\omega_{DE}})\frac{\theta_{DE}}{\bar{H}}\nonumber\\
&&+\frac{\bar{H}\theta_{DE}}{(1+\mathrm{z})^{2}\mathrm{k^{2}}}\biggl[9(1-\mathrm{{\omega}_{DE}^{2}})-3(1+\mathrm{z})\frac{\mathrm{d}{\omega}_{DE}}
{\mathrm{d}z}\nonumber\\
&&+3\mathrm{R}{\bar{\rm I}}_{\rm Q}(1-\mathrm{\omega_{DE}})\biggr]+3(1+\mathrm{\omega_{DE}})\frac{\mathrm{d}{\psi}}{\mathrm{d}z}+\nonumber\\
&&+\frac{\mathrm{R}{\bar{\rm I}}_{\rm Q}\phi}{(1+\mathrm{z})}+\frac{\mathrm{R}{\bar{\rm I}}_{\rm Q}}{(1+\mathrm{z})}\delta_{DM},\\
\label{contrastDM}
\frac{\mathrm{d}{\delta}_{DM}}{\mathrm{d}z}&=&\frac{\theta_{DM}}{\bar{H}}-\frac{{\bar{\rm I}}_{\rm Q}\phi}{(1+\mathrm{z})}+3\frac{\mathrm{d}{\psi}}
{\mathrm{d}z}\,,\\
\label{contrastb}
\frac{\mathrm{d}{\delta}_{b}}{\mathrm{d}z}&=&\frac{\theta_{b}}{\bar{H}}+3\frac{\mathrm{d}{\psi}}{\mathrm{d}z},\\
\label{eulerDE1}
\frac{\mathrm{d}{\theta}_{DE}}{\mathrm{d}z}&=&-\frac{2{{\theta}_{DE}}}{(1+\mathrm{z})}\left(1+\frac{\mathrm{R}{\bar{\rm I}}_{\rm Q}}
{(1+\mathrm{\omega_{DE}})}\right)-\frac{\mathrm{k^{2}}{\delta}_{DE}}{\bar{H}(1+\mathrm{\omega_{DE}})}\nonumber\\
&&-\frac{\mathrm{k^{2}}\phi}{\bar{H}}+\frac{\mathrm{R}{\bar{\rm I}}_{\rm Q}{\theta}_{DM}}{(1+\mathrm{z})(1+\mathrm{\omega_{DE}})},\\
\label{eulerDM1} 
\frac{\mathrm{d}{\theta}_{DM}}{\mathrm{d}z}&=&\frac{\theta_{DM}}{(1+\mathrm{z})}-\frac{k^{2}\phi}{\bar{H}},\\
\label{eulerb1}
\frac{\mathrm{d}{\theta}_{b}}{\mathrm{d}z}&=&\frac{\theta_{b}}{(1+\mathrm{z})}-\frac{k^{2}\phi}{\bar{H}}.
\end{eqnarray}
Furthermore, the relativistic Poisson equation is given by
\begin{eqnarray}\label{poisson}
 \mathrm{k}^{2}\phi&=&\frac{3\bar{H}^{2}}{(1+\mathrm{z})}\frac{\mathrm{d}{\psi}}{\mathrm{d}z}-\frac{3{\bar{H}}^{2}\phi}{(1+\mathrm{z})^{2}}
 -\frac{4\pi G}{(1+\mathrm{z})^{2}}\times\nonumber\\
 &&\biggl(\bar{\rho}_{b}\delta_{b}+\bar{\rho}_{DM}\delta_{DM}+\bar{\rho}_{DE}\delta_{DE}+\bar{\rho}_{r}\delta_{r}\biggr).
\end{eqnarray}
\subsection{Structure formation}\label{growth factor}
In the Newtonian limit ($\delta\ll1$) and at sub-horizon scales ($\bar{H}^{2}\ll\mathrm{k}^{2}$), we assume that DE fluid does not 
contribute in clustering of matter and therefore we could take $\delta_{DM}\gg\delta_{DE}\approx 0$. Besides, for simplicity, we also consider that 
the gravitational potentials $\phi$ and $\psi$, satisfy $\delta_{DM}\gg \phi=\psi$ and $\phi'=\psi'\approx 0$.\\
Since we are only interested in showing the effects of $\bar{Q}$ and $\mathrm{{\omega}_{DE}}$ on the evolution of $\delta_{DM}$ during the matter 
dominated era, rather than making accurate calculations. Again, for simplicity, we can ignore the contribution of the radiation in our estimations. 
Due to the arguments above discussed and combining Eqs. (\ref{contrastDM}), (\ref{eulerDM1}) and (\ref{poisson}), we obtain 
\begin{eqnarray}\label{secondDM1} 
\frac{\mathrm{d}^{2}{\delta}_{DM}}{\mathrm{d}{\mathrm{z}}^{2}}&=&-\frac{(1+3\mathrm{\omega_{DE}}\mathrm{\bar{\Omega}_{DE}}+\mathrm{\bar{\Omega}_{r}})}{2(1+\mathrm{z})}
\frac{\mathrm{d}{\delta}_{DM}}{\mathrm{dz}}\nonumber\\
&&+\frac{3}{2(1+\mathrm{z})^{2}}\biggl(\mathrm{\bar{\Omega}_{DM}}\delta_{DM}+\mathrm{\bar{\Omega}_{b}}\delta_{b}\biggr).
\end{eqnarray}
A similar equation can be obtained for $\delta_{b}$.\\
Next, we define the growth factor of linear matter perturbations as
\begin{equation}\label{definitionf} 
\mathrm{f}\equiv \frac{{d}{\ln}\delta_{\mathrm M}}{d{\ln \mathrm{\bf a}}}\,,\quad  {{\delta}_{\mathrm M}=\mathrm{\Omega_{DM}}\delta_{DM}+
\mathrm{\Omega_{b}}\delta_{b}},
\end{equation}
where ${\delta}_{\mathrm M}$ is the normalized matter density perturbations.\\ 
Via the above definition and by considering that $\delta_{b}\ll\delta_{DM}$, Eq. (\ref{secondDM1}) can be re-expressed in terms of the redshift for 
the case $Q^{\mu}_{A}\parallel U^{\mu}_{DM}$, as
\begin{equation}
\label{factor1}
\frac{d\mathrm{f}}{d\mathrm{z}}=\frac{1}{\mathrm{(1+z)}}\biggl[\mathrm{f}^{2}+\frac{\mathrm{f}}{2}\left(1-3\mathrm{\omega_{DE}}
\mathrm{\bar{\Omega}_{DE}}-\mathrm{\bar{\Omega}_{r}}\right)-\frac{3}{2}\mathrm{\bar{\Omega}_{DM}}\biggr].
\end{equation}
An analytical solution to Eq. (\ref{factor1}) is very complicated to obtain, and we need to use numerical methods. For this reason, 
it is most suitable to approach $\mathrm{f}$ in the form
\begin{equation}\label{ansatzf} 
\mathrm{f}={\bar{\Omega}_{M}}^{\gamma}\,, 
\end{equation}
where $\gamma$ is the growth index of the linear matter fluctuations, and in general is a function of the redshift or scale factor. 
Hence, Eq. (\ref{secondDM1}) can be solved numerically taking into account the conditions at $\mathrm{z}=0$:  
$\mathrm{\delta_{M,0}|=1}$ and ${\mathrm{d{\delta}_{M}}}/{\mathrm{dz}}|_{\mathrm{z}=0}=-\mathrm{\Omega_{M,0}}^{\mathrm{\gamma_{0}}}$, 
where $\Omega_{M,0}$ and $\gamma_{0}$ are the values today, and $\mathrm{\Omega_{M,0}}=\mathrm{\Omega_{DM,0}+\Omega_{b,0}}$.\\
Similarly, by considering the condition $\mathrm{f_{0}}=\mathrm{{\Omega_{M,0}}^{\gamma_{0}}}$, Eq. (\ref{factor1}) can also be solved numerically.\\ 
On the other hand, the root-mean-square amplitude of matter density perturbations within a sphere of radius $\mathrm{8\,Mpc}h^{-1}$ is denoted as 
$\mathrm{\sigma_{8}(z)}$ and its evolution is represented by
\begin{equation}\label{Sigma8}
\mathrm{\sigma_{8}(z)}=\delta_{M}(\mathrm{z})\mathrm{\sigma_{8,0}},
\end{equation}
where $\mathrm{\sigma_{8,0}}$ is the normalizations to unity of $\mathrm{\sigma_{8}(z)}$ today. Thus, the functions $\mathrm{f}$ y $\mathrm{\sigma_{8}}$ can be 
combined to obtain $\mathrm{f\sigma_{8}}$ at different redshifts. From here, we obtain 
\begin{equation}\label{fs8}
\mathrm{f(z)\sigma_{8}(z)}=\mathrm{f(z)}\delta(\mathrm{z})\mathrm{\sigma_{8,0}}.
\end{equation}
\subsection{Linear matter power spectrum.}\label{spectrum1}
The linear matter power spectrum $\mathrm{P(k,z)}$ is \cite{Hu1998,Dodelson2003}
\begin{equation}\label{pk}
\mathrm{P(k,z)}=2\pi^{2}\mathrm{H_{0}}^{-(3+n_{s})}\delta_{H}^{2}k^{n_{s}}T^{2}(\mathrm{k})\delta_{M}^{2}(\mathrm{z}),
\end{equation}
where ${T(\mathrm{k})}$ is the transfer function, $n_{s}$ is the scalar spectral index of the primordial fluctuation spectrum, $\mathrm{k}$ is the wavenumber and 
$\delta_{H}$ is defined \cite{Hu1998,Dodelson2003} by
\begin{eqnarray}\label{dh}
\delta_{H}\approx 1.94\times 10^{-5}\mathrm{\Omega_{M,0}}^{-0.785-0.05\mathrm{\ln\Omega_{M,0}}}e^{-0.95(n_{s}-1)}.\,\,
\end{eqnarray}
In this work, we adopt the fitting formula proposed in \cite{Hu1998,Dodelson2003} that approximates the full transfer function as the sum of the 
baryon and cold DM contribution on all scales 
\begin{equation}\label{fulltransfer}
T(\mathrm{k})=\mathrm{\frac{\Omega_{b,0}}{\Omega_{M,0}}}T_{b}(\mathrm{k})+\mathrm{\frac{\Omega_{DM,0}}{\Omega_{M,0}}}T_{DM}(\mathrm{k}). 
\end{equation}
Here, $T_{b}(k)$ is the baryon transfer function defined as
\begin{eqnarray}\label{transferb}
T_{b}(\mathrm{k})&=&\biggl[\frac{1}{1+(\frac{\mathrm{k}r_{d}}{5.2})^{2}}\frac{\ln(e+1.8q)}{\ln(e+1.8q)+C_{0}q^{2}}+\nonumber\\
&&\frac{\alpha_{b}e^{-(\mathrm{k/k_{silk}})^{1.4}}}{1+(\beta_{b}/\mathrm{k}r_{d})^{3}}\biggr]\frac{sin(\mathrm{k}\tilde{r})}{\mathrm{k}\tilde{r}},\nonumber\\
q&=&\frac{\mathrm{k}}{13.41\mathrm{k_{eq}}},\quad \alpha_{b}=\frac{2.07\mathrm{k_{eq}}r_{d}G}{(1+R_{d})^{0.75}}\frac{\mathrm{a_{*}}}{\mathrm{a_{eq}}},\nonumber\\
r_{d}&=&\frac{2}{3\mathrm{k_{eq}}}\sqrt\frac{6}{R_{eq}}\ln\biggl[\frac{\sqrt{R_{d}+R_{eq}}+\sqrt{1+R_{d}}}{1+\sqrt{R_{eq}}}\biggr],\nonumber\\
\tilde{r}&=&\frac{r_{d}}{[1+(\frac{\beta_{node}}{\mathrm{k}r_{d}})^{3}]^{1/3}}\quad \beta_{node}=8.41(\mathrm{\Omega_{m0}}h^{2})^{0.435},\nonumber\\
\mathrm{k_{silk}}&=&1.6(\mathrm{\Omega_{b0}}h^{2})^{0.52}(\mathrm{\Omega_{m0}}h^{2})^{0.73}\biggl[1+\frac{1}{(10.4\mathrm{\Omega_{m0}}h^{2})^{0.95}}\biggr],\nonumber\\
\mathrm{\beta_{b}}&=&0.5+\mathrm{\frac{\Omega_{b,0}}{\Omega_{m0}}+(3-\frac{2\Omega_{b0}}{\Omega_{m0}})}
\sqrt{(17.2\mathrm{\Omega_{m0}}h^{2})^{2}+1},\,
\end{eqnarray}
where $R_{d}$ is the ratio of the baryon to photon energy density at the drag epoch, $\mathrm{k_{eq}}$ is the wave-number at the equality epoch 
radiation-matter, $r_{d}(z_{d})$ is the sound horizon at the drag epoch, $G$ is a factor of suppression, $\mathrm{a_{*}}$ represents the scale factor at 
the recombination epoch, $\mathrm{a_{eq}}$ represents the scale factor at the equality epoch radiation-matter and $\mathrm{k_{silk}}$ represents the 
Silk damping scale \cite{Hu1998,Dodelson2003}.\\
Similarly, the cold DM transfer function, $T_{DM}(\mathrm{k})$, is defined as \cite{Hu1998,Dodelson2003}.
\begin{eqnarray}\label{transferdm}
T_{DM}(\mathrm{k})&=&\frac{{L_{0}}}{{L_{0}+C_{0}q^{2}}}\,,\quad {L_{0}}={\ln(5.436+1.8q)},\nonumber\\
{C_{0}}&=&14.2+\frac{731}{1+62.5{q}},\quad {q}=\frac{\mathrm{k}}{h\Gamma},
\end{eqnarray}
and the shape parameter $\Gamma$, is given by \cite{Hu1998,Dodelson2003}
\begin{eqnarray}\label{shape}
\Gamma\equiv\mathrm{\Omega_{M,0}}h\left(\zeta+\frac{1-\zeta}{[1+0.43\mathrm{k}r_{s}(z_{d})]^{4}}\right),\,\,\upsilon=\mathrm{\frac{\Omega_{b,0}}{\Omega_{M,0}}},\quad\\
\zeta=1-0.328\ln(431\mathrm{\Omega_{M,0}}h^{2})\upsilon+0.38\ln(22.3\mathrm{\Omega_{M,0}}h^{2})\upsilon^{2}.\nonumber
\end{eqnarray}
\subsection{CMB temperature power spectrum.}\label{spectrum2}
From Eqs. (\ref{pk})-(\ref{transferdm}), using the results found in \cite{Dodelson2003} and the Limber approximation \cite{Limber1953}, we have 
built numerically the CMB temperature power spectrum today as
\begin{eqnarray}\label{clpower}
C_{l}\,\alpha\,\frac{B_{1}}{l}cos^{2}\left(\frac{lr_{s}(z_{*})}{D_{A}}\right)+\frac{B_{2}}{l}cos\left(\frac{lr_{s}(z_{*})}{D_{A}}\right)T(l,D_{A})\nonumber\\
+\frac{B_{3}}{l}T^{2}(l,D_{A})+B_{4}sin^{2}\left(\frac{lr_{s}(z_{*})}{D_{A}}\right)+...,\qquad
\end{eqnarray}
where the respective coefficients $B_{1}$, $B_{2}$, $B_{3}$ and $B_{4}$ are functions of $\mathrm{H_{0}}$, $D_{A}(z_{*})$, $\mathrm{\Omega_{M,0}}$, 
$\delta_{H}$, $\delta_{M}$, $R_{*}$, $\mathrm{k}$, $\mathrm{k_{eq}}$, $\mathrm{k_{silk}}$ and $\tau$. Here, $D_{A}(z_{*})$ represents the angular 
diameter distance at the recombination epoch, see Eq. (\ref{DA}), ${R_{*}}$ is the ratio of the baryon to photon energy density at the recombination epoch and ${r_{s}(z_{*})}$ is the sound 
horizon at the recombination epoch, see Eq. (\ref{Redshift_decoupling}), and $\tau$ represents the optical depth.
\section{Constraint method and observational data}
\subsection{Constraint method}
In general, to constrain the parameter space we build all the necessary codes in the c++ language and use the MCMC method to calculate the best-fit 
parameters of the $\Lambda$CDM, $\omega$DE and IDE models, respectively, and their respective parameter space P (main parameters), are given by
\begin{eqnarray*}
\mathrm{P_{1}}&\equiv& \mathrm{\biggl(\Omega_{DM,0},\mathrm{H_{0}},\alpha,\beta,M,dM,\gamma_{0},\sigma_{80}\biggr)},\nonumber\\
\mathrm{P_{2}}&\equiv& \mathrm{\biggl(\omega_{0},\omega_{1},\omega_{2},\Omega_{DM,0},\mathrm{H_{0}},\alpha,\beta,M,dM,\gamma_{0},\sigma_{80}\biggr)},\nonumber\\
\mathrm{P_{3}}&\equiv& \mathrm{\biggl({\lambda}_{0},{\lambda}_{1},{\lambda}_{2},\omega_{0},\omega_{1},\omega_{2},\Omega_{DM,0},\mathrm{H_{0}},\alpha,\beta,M,dM,
\gamma_{0},\sigma_{80}\biggr)},
\end{eqnarray*}
where $\mathrm{\Omega_{DM,0}}$ and $\mathrm{H_{0}}$ are the DM energy density and the Hubble parameter today, $\mathrm{{\omega}_{0}}$, $\mathrm{{\omega}_{1}}$ and $\mathrm{{\omega}_{2}}$ are dimensionless parameters related to $\mathrm{\omega_{DE}}$. 
Similarly, $\mathrm{{\lambda}_{0}}$, $\mathrm{{\lambda}_{1}}$ and $\mathrm{{\lambda}_{2}}$ are dimensionless constants linked to $\bar{Q}$. 
The nuisance parameters $\mathrm{\alpha}$, $\mathrm{\beta}$, $\mathrm{M}$ and $\mathrm{dM}$ are connected with the global properties of the 
Supernovas (type Ia), $\mathrm{\gamma_{0}}$ and $\mathrm{\sigma_{80}}$ are the values of $\mathrm{\gamma}$ and $\mathrm{\sigma_{8}}$ today, 
respectively. The pivot scale of the initial scalar power spectrum $\mathrm{k_{s,0}=0.045Mpc^{-1}}$ is assumed. Besides, the constant priors for the 
model parameters are shown in Table \ref{Priors}. We have also fixed $\mathrm{\Omega_{r,0}}=\mathrm{\Omega_{\gamma,0}}(1+0.2271N_{eff})$, where 
$N_{eff}$ represents the effective number of neutrino species $N_{eff}=3.04\pm0.18$ and $\mathrm{\Omega_{\gamma,0}}=2.469\times10^{-5}h^{-2}$ were 
chosen from Table $4$ in \cite{Planck2015}. Similarly, the values of $\mathrm{\Omega_{b,0}}=0.02230h^{-2}$ and the Gaussian prior on 
$n_{s}=0.9667 \pm 0.0040$ were also taken from Table $4$ in \cite{Planck2015}.\\
Furthermore, the dimensionless parameters such as the ratio of the sound horizon and angular diameter distance, $\Theta_{s}$ (multiplied by 100), 
together with the optical depth $\tau$ and the amplitude of the initial power spectrum $A_{s}$, are derived from the parameter space P.\\
In order to have access to the distribution of $\mathrm{P}$, we calculate the overall likelihood 
$\mathcal{L}\, \alpha\, {{\mathrm e}}^{-{{\chi}}^{2}/2}$, where ${\rm {\bf{\chi}}^{2}}$ is 
\begin{equation}\label{TotalChi}
{{{\rm {\bf{\chi}}^{2}}}}={{{\rm{\bf{\chi}}^{2}}}}_{\bf JLA}+{{{\rm {\bf {\chi}}^{2}}}}_{\bf{RSD}}+{{{\rm {\bf {\chi}}^{2}}}}_{\bf BAO}+{{{\rm {\bf {\chi}}^{2}}}}_{\bf CMB}+
{{{\rm {\bf {\chi}}^{2}}}}_{\bf H}\,.
\end{equation}
\subsection{Observational data}
To test the viability of our model and set constraints on $\mathrm{P}$, we use the following data sets:
\subsubsection{Join Analysis Luminous data set (JLA).} \label{SNIa}
The Supernovae (SNe Ia) data sample used in this work is the Join Analysis Luminous data set (JLA) \cite{Conley2011,Jonsson2010,Betoule2014} composed by 
$740$ SNe with high-quality light curves. Here, JLA data include samples from $\mathrm{z}<0.1$ to $0.2<\mathrm{z}<1.0$.\\
The observed distance modulus is modeled by \cite{Conley2011,Jonsson2010,Betoule2014}
\begin{equation}\label{muJLA}
{\mu}^{JLA}_{i}={m}^{*}_{B,i}+\mathrm{\alpha}{x_{1,i}}-\mathrm{\beta}C_{i}-M_{B},\quad 1\leq i\leq740,
\end{equation}
where and the parameters ${m}^{*}_{B}$, $x_{1}$ and $C$ describe the intrinsic variability in the luminosity of the SNe. Furthermore, the nuisance parameters 
$\mathrm{\alpha}$, $\mathrm{\beta}$, $\mathrm{M}$ and $\mathrm{dM}$ characterize the global properties of the light-curves of the SNe and are estimated simultaneously with the cosmological parameters 
of interest. Then, we defined $M_{B}$
\begin{equation}\label{Mb}
M_{B}=\left\{\begin{array}{cl}
\mathrm{M},\, &\mbox{if}\,\,\,\rm{M_{stellar}}\,\,\,<10^{10}\rm{M_{\bigodot}},\\
\mathrm{M}+\mathrm{dM},\, &\mbox{if}\,\,\,\rm{otherwise}\,,           
\end{array}\right.
\end{equation}
where $\rm{M_{stellar}}$ is the host galaxy stellar mass, and $\rm{M_{\bigodot}}$ is the solar mass.\\
On the other hand, the theoretical distance modulus is 
\begin{equation}\label{mus}
{\mu}^{\rm{th}}(\mathrm{z},\mathbf{X}) \equiv 5{\log}_{10}\left[\frac{{D_{L}}(\mathrm{z},\mathbf{X})}{\rm{Mpc}}\right]+25,
\end{equation}
where ``$\rm{th}$'' denotes the theoretical prediction for a SNe at $\mathrm{z}$. The luminosity distance ${D_{L}}(\mathrm{z},\mathbf{X})$, is defined as
\begin{equation}\label{luminosity_distance1}
{D}_{L}(z_{hel},z_{CMB},\mathbf{X})=(1+z_{hel})c \int_{0}^{z_{CMB}}\frac{dz'}{\mathrm{H}(z',\mathbf{X})},
\end{equation}
where $z_{hel}$ is the heliocentric redshift, $z_{CMB}$ is the CMB rest-frame redshift, ``$c=2.9999\times10^{5}km/s$'' is the speed of the light and $\mathbf{X}$ represents 
the model parameters. Thus, we rewrite ${\mu}^{{\rm th}}(\mathrm{z},\mathbf{X})$ as
\begin{eqnarray}\label{mus}
{\mu}^{{\rm th}}(z_{hel},z_{CMB},\mathbf{X})&=&5\log_{10}\biggl[(1+z_{hel}\int_{0}^{z_{CMB}}\frac{dz'}{E(z',\mathbf{X})})\biggr]\nonumber\\
&&+52.385606-5\log_{10}(\mathrm{H_{0}}).
\end{eqnarray}
Then, the ${\chi}^{2}$ distribution function for the JLA data is
\begin{equation}\label{X2JLA}
{\chi}_{\bf JLA}^{2}(\mathbf{X})=\left({\Delta{\mu}}_{i}\right)^{t}\left(C^{-1}_{\bf Betoule}\right)_{ij}\left({\Delta{\mu}}_{j}\right),
\end{equation}
where ${\Delta \mu}_{i}={\mu}^{th}_{i}(\mathbf{X})-{\mu}^{JLA}_{i}$ is a column vector and $C^{-1}_{\bf Betoule}$ is the $740\times740$ covariance 
matrix \cite{Betoule2014}.
\subsubsection{Redshift Space Distortion (RSD) data}\label{RSDdata}
Represent a compilation of measurements of the quantity $\mathrm{f\sigma_{8}}$ at different 
redshifts, and obtained in a model independent way. These data are apparent anisotropies of the galaxy distribution in redshift space 
due to the differences of the estimates between the redshift observed distances and true distances. Here, $\mathrm{f}$ is combined with the 
root-mean-square amplitude of matter within a sphere of radius $\mathrm{8\,Mpc h^{-1}}$, $\mathrm{\sigma_{8}(z)}$, in a single quantity. 
These data were derived from the following galaxy surveys: Pscz, 2dFVVDS, 6dF, 2MASS, BOSS and WiggleZgalaxy, respectively and collected by 
Mehrabi (see Table in \cite{Mehrabi2015}). Then, the standard $\chi^2$ for this data set is 
given as \cite{Mehrabi2015}
\begin{equation}\label{X2RSD}
{\chi}^{2}_{RSD}(\mathbf{X}) \equiv \sum_{i=1}^{18}\frac{\left[\mathrm{f\sigma_{8}}^{\mathrm{th}}(\mathbf{X},\mathrm{z_{i}})-\mathrm{f\sigma_{8}}^{\mathrm{obs}}(\mathrm{z_{i}})\right]^{2}}{{\sigma}^{2}(\mathrm{z_{i}})},
\end{equation}
where $\sigma(\mathrm{z_{i}})$ is the observed $1\sigma$ error, $\mathrm{f\sigma_{8}}^{\rm{th}}$ and $\mathrm{f\sigma_{8}}^{\mathrm{obs}}$ 
denote the theoretical and observational data, respectively.
\subsubsection{BAO data sets}\label{BAO}
$\bullet$ ${\mathbf{\it {BAO\,I\,data}}}$: Here, we use a compilation of measurements of the distance ratios $\mathrm{d_{z}}$ at different redshifts and 
obtained from different surveys 
\cite{Hinshaw2013,Beutler2011,Ross2015,Percival2010,Kazin2010,Padmanabhan2012,Chuang2013a,Chuang2013b,Anderson2014a,Kazin2014,Debulac2015,FontRibera2014},
listed in Table \ref{tableBAOI}. To encode the visual distortion of a spherical object due to the non-Euclidianity of a FRW spacetime, the 
authors \cite{Percival2010,Eisenstein1998} constructed a distance ratio $D_{v}(\mathrm{z})$
\begin{equation}\label{Dv}
D_{v}(\mathrm{z},\mathbf{X})\equiv\frac{1}{\mathrm{H_{0}}}\left[(1+\mathrm{z})^{2}{D_{A}}^{2}(\mathrm{z})\frac{c\mathrm{z}}{E(\mathrm{z})}\right]^{1/3},
\end{equation}
where $D_{A}(\mathrm{z})$ is the angular diameter distance given by
\begin{eqnarray} \label{DA}
D_{A}(\mathrm{z},\mathbf{X}) &\equiv& c{\int}^{\mathrm{z}}_{0}\frac{dz'}{\mathrm{H}(z',\mathbf{X})}.
\end{eqnarray}
The comoving sound horizon size is defined by
\begin{equation} \label{hsd} 
r_{s}(\mathrm{a})\equiv c\int^{\mathrm{\bf a}}_{0}\frac{c_{s}(a')da'}{{a'}^{2}H(a')},
\end{equation}
being $c_{s}(a)$ the sound speed of the photon-baryon fluid
\begin{equation} \label{vsd}
c_{s}^{2}(\mathrm{\bf a}) \equiv \frac{\delta P}{\delta \rho}=\frac{1}{3}\left[\frac{1}{1+(3\mathrm{\Omega_{b}}/4\mathrm{\Omega_{r}})\mathrm{\bf a}}\right].
\end{equation}
Considering Eqs. (\ref{hsd}) and (\ref{vsd}) in terms of $\mathrm{z}$, we have   
\begin{equation}\label{rs}
r_{s}(\mathrm{z})=\frac{c}{\sqrt 3}{\int}^{1/(1+\mathrm{z})}_{0}\frac{\mathrm{da}}{\mathrm{{a}^{2}}\mathrm{H(a)}\sqrt{1+(3\mathrm{\Omega_{b,0}}/4\mathrm{\Omega_{\gamma,0}})\mathrm{\bf a}}}.
\end{equation}
The epoch in which the baryons were released from photons is denoted as, $z_{d}$, and can be determined by \cite{Eisenstein1998}:
\begin{equation}\label{zd}
\mathrm{z_{d}}=\frac{1291(\mathrm{\Omega_{M,0}}h^{2})^{0.251}}{1+0.659(\mathrm{\Omega_{M,0}}h^{2})^{0.828}}\left(1+b_{1}(\mathrm{\Omega_{b,0}}h^{2})^{b_{2}}\right),
\end{equation}
where $\mathrm{\Omega_{M,0}}=\mathrm{\Omega_{DM,0}+\Omega_{b,0}}$, and
\begin{eqnarray*} 
b_{1}&=&0.313(\mathrm{\Omega_{M,0}}h^{2})^{-0.419}\left[1+ 0.607(\mathrm{\Omega_{M,0}}h^{2})^{0.674}\right],\\
b_{2}&=&0.238(\mathrm{\Omega_{M,0}}h^{2})^{0.223}.
\end{eqnarray*}
The peak position of the BAO depends on the distance radios $\mathrm{d_{z}}$ at different $\mathrm{z}$, which are listed in Table \ref{tableBAOI}.\\
\begin{equation} \label{dvalues}
\mathrm{d_{z}(\mathbf{X})}=\frac{r_{s}(\mathrm{z_{d}})}{D_{V}(\mathrm{z},\mathbf{X})},
\end{equation}
where $r_{s}(\mathrm{z_{d}},\mathbf{X})$ is the comoving sound horizon size at the baryon drag epoch. From Table \ref{tableBAOI}, the $\chi^{2}$ becomes 
\begin{equation}\label{X2BAOI}
\chi_{\bf{{BAO}}\,\rm{I}}^{2}(\mathbf{X})=\sum_{i=1}^{17}\left(\frac{d_{z}^{\mathrm{th}}(\mathbf{X},\mathrm{z_{i}})-d_{z}^{\mathrm{obs}}(\mathbf{X},\mathrm{z_{i}})}{\sigma(\mathbf{X},\mathrm{z_{i}})}\right)^{2}.
\end{equation}
$\bullet$ ${\mathbf{\it {BAO\,II\,data}}}$: From BOSS DR $9$ CMASS sample, Chuang \cite{Chuang2013b} analyzed the shape of the monopole and quadrupole 
from the two-dimensional two-points correlation function $2$d$2$pCF of galaxies and measured simultaneously $\mathrm{H}$, $D_{A}$, 
$\Omega_{m}h^{2}$ and $\mathrm{f\sigma_{8}}$ at the effective redshift $\mathrm{z}=0.57$, and then, defined 
${\Delta A}_{i}=A^{\mathrm{th}}_{i}(\mathbf{X})-A^{\mathrm{obs}}_{i}$ as a column vector
\begin{equation}\label{bossdr9}
{\Delta A}_{i}=\left(\begin{array}{rl}
\mathrm{H(0.57)}-87.6\\
D_{A}(0.57)-1396\\
\Omega_{m}h^{2}(0.57)-0.126\\
\mathrm{f(0.57)\sigma_{8}(0.57)}-0.428\\
\end{array}\right),
\end{equation}
Then, the $\chi^{2}$ function for the BAO $\rm{II}$ data is given by 
\begin{equation}\label{X2BAOII}
\chi_{\bf BAO\,\rm{II}}^{2}(\mathbf{X})=\left({\Delta A}_{i}\right)^{t}\left(C^{-1}_{\bf BAO\,\rm{II}}\right)_{ij}\left({\Delta A}_{j}\right),
\end{equation}
where ``t'' denotes its transpose and the covariance matrix is listed in Eq. ($26$) of \cite{Chuang2013b}
\begin{equation}\label{BAOII}
C^{-1}_{\bf{\rm{BAO}}\,\bf{\rm{II}}}=\left(\begin{array}{lccr}
 +0.03850 \,\,-0.0011410 \,-13.53 \,-1.2710\\
 -0.001141 +0.0008662   \,+3.354 \,-0.3059\\
 -13.530  \,\,\,\,\,+3.3540\quad\,\,\,\,\,+19370\,-770.0\\
 -1.2710  \,\,\,\,\,-0.30590 \quad\,-770.0\,\,+411.3
\end{array} \right).
\end{equation}
$\bullet$ ${\mathbf{\it {BAO\,III\,data}}}$: Using SDSS DR $7$ sample Hemantha \cite{Hemantha2014}, proposed a new method to constrain 
$\mathrm{\bar{H}}$ and $D_{A}$ simultaneously from the two-dimensional matter power spectrum $2$dMPS without assuming a DE model or a flat universe. 
They defined a column vector ${\Delta B}_{i}=B^{\mathrm{th}}_{i}(\mathbf{X})-B^{\mathrm{obs}}_{i}$ as
\begin{equation}
B^{\mathrm{th}}_{i}(\mathbf{X})-B^{\mathrm{obs}}_{i}=\left(\begin{array}{rl}                               
 \mathrm{H(0.35,\mathbf{X})}-81.3\\
 D_{A}(0.35,\mathbf{X})-1037.0\\
 \Omega_{M}h^{2}(0.35,\mathbf{X})-0.1268\\
\end{array}\right).
\end{equation}
The covariance matrix for the set of parameters was
\begin{equation}\label{BAOIII}
C^{-1}_{\bf BAO\,\rm{III}}=\left(\begin{array}{lcr}           
+0.00007225 \, -0.169606 \,+0.01594328\\
-0.1696090  \,\,\,+1936.0\,\,\quad+67.030480 \\
+0.01594328 \, +67.03048\,+14.440\\
\end{array} \right).
\end{equation}
The $\chi^{2}$ function for these data can be written as  
\begin{equation}\label{X2BAOIII}
\chi_{\bf{\rm{BAO}}\,\bf{\rm{III}}}^{2}(\mathbf{X})=\left({\Delta B}_{i}\right)^{t}\left(C^{-1}_{\bf BAO \rm{III}}\right)_{ij}\left({\Delta B}_{j}\right),
\end{equation}
where ``t'' denotes its transpose.\\
$\bullet$ ${\mathbf{\it {BAO\,IV\,data}}}$:
This sample considers the Alcock-Paczynski test \cite{Alcock1979} to constrain the cosmological models and break the degeneracy between $D_{A}$ 
and $\mathrm{\bar{H}}$ \cite{Seo2008}. This signal can be defined through the $\mathrm{AP}$ distortion parameter 
$F_{AP}(\mathrm{z})=(1+\mathrm{z})D_{A}(\mathrm{z})\left(\mathrm{H(z)}/c\right)$. 
In this sample has been convenient to define the joint measurements of $\mathrm{d_{z}(z_{eff}})$, $F_{AP}(\mathrm{z_{eff}})$ and 
$\mathrm{f(z_{eff})\sigma_{8}(z_{eff})}$ in a only vector evaluated at the effective redshift $\mathrm{z_{eff}}=0.57$ 
\cite{Battye2015,Samushia2014,Anderson2014a}.
Here, it is convenient to define the joint measurements of $\mathrm{d_{z}(z_{eff}})$, $F_{AP}(\mathrm{z_{eff}})$ and 
$\mathrm{f(z_{eff})\sigma_{8}(z_{eff})}$ in a vector $V$ evaluated at the effective redshift $\mathrm{z_{eff}}=0.57$ 
\cite{Battye2015,Samushia2014,Anderson2014a}
\begin{equation}\label{bossdr11}
{\Delta V}_{i}=V^{\mathrm{th}}_{i}(\mathbf{X})-V^{\mathrm{obs}}_{i}=\left(\begin{array}{rl}
\mathrm{d_{z}(z_{eff})}-13.880\\
F_{AP}(\mathrm{z_{eff}})-0.683\\
\mathrm{f(z_{eff})\sigma_{8}(z_{eff})}-0.422\\
\end{array}\right),
\end{equation}
The $\chi^{2}$ function for this data set is fixed as
\begin{equation}\label{X2BAOIV}
\chi_{\bf BAO\,\rm{IV}}^{2}(\mathbf{X})=\left({\Delta V}_{i}\right)^{t}\left(C^{-1}_{\bf BAO\,\rm{IV}}\right)_{ij}\left({\Delta V}_{j}\right)\,,
\end{equation}
where the covariance matrix is listed in Eq. ($1.3$) of \cite{Battye2015}
\begin{equation}\label{BAOIV}
C^{-1}_{\bf{\rm{BAO}}\,\bf{\rm{IV}}}=\left(\begin{array}{lcr}           
+31.032  \,+77.773 \,-16.796\\
+77.773 \, +2687.7 \,-1475.9\\
-16.796 \, -1475.9 \,+1323.0\\
\end{array} \right).
\end{equation}
Considering Eqs. (\ref{X2BAOI}), (\ref{X2BAOII}), (\ref{X2BAOIII}) and (\ref{X2BAOIV}), we can construct the total $\chi_{\bf BAO}^{2}$ 
for all the BAO data, as
\begin{equation}\label{X2Total}
 {{{\rm{\bf{\chi}}^{2}}}}_{\bf BAO}=\chi_{\bf{BAO\,\rm{I}}}^{2}+\chi_{\bf{BAO\,\rm{II}}}^{2}+\chi_{\bf{BAO\,\rm{III}}}^{2}+\chi_{\bf{BAO\,\rm{IV}}}^{2}\,.
\end{equation}
\subsubsection{CMB data} \label{CMB}
We use the Planck distance priors data extracted from Planck $2015$ results XIII Cosmological parameters \cite{Planck2015}. From here, we have 
obtained the Shift parameter ${\tilde{R}\mathrm{(z_{*})}}$, the angular scale for the sound horizon at recombination epoch, $l_{A}(\mathrm{z_{*}})$, where 
$\mathrm{z_{*}}$ represents the redshift at recombination epoch \cite{Planck2015,Neveu2016}.\\
Hence, the shift parameter $\tilde{R}$ is defined by \cite{Bond-Tegmark1997}
\begin{equation}\label{Shiftparameter}
\tilde{R}(z_{*},\mathbf{X})\equiv\sqrt{\Omega_{M,0}}{\int}^{z_{*}}_{0}\frac{d\tilde{y}}{E(\tilde{y})},
\end{equation}
where $E(\tilde{y})$ is given by Eq. (\ref{hubble}). The redshift $z_{*}$ is obtained using \cite{Hu-Sugiyama1996}
\begin{equation}\label{Redshift_decoupling}
{z}_{*}=1048\biggl[1+0.00124(\mathrm{{\Omega}_{b,0}}h^{2})^{-0.738}\biggr]\biggl[1+{g}_{1}(\mathrm{{\Omega}_{M,0}}h^{2})^{{g}_{2}}\biggr]\;,\;\;
\end{equation}
where 
\begin{equation}\label{g1g2}
g_{1}=\frac{0.0783(\mathrm{\Omega_{b,0}}h^{2})^{-0.238}}{1+39.5(\mathrm{\Omega_{b,0}}h^{2})^{0.763}},\hspace{0.3cm}
g_{2}=\frac{0.560}{1+21.1(\mathrm{\Omega_{b,0}}h^{2})^{1.81}}.
\end{equation}
The angular scale $l_{A}$ for the sound horizon at recombination epoch is 
\begin{equation}\label{Acoustic_scale}
l_{A}(\mathbf{X})\equiv\frac{\pi D_{A}(z_{*},\mathbf{X})}{r_{s}(z_{*},\mathbf{X})},\hspace{1cm}
\end{equation}
where $r_{s}(z_{*},\mathbf{X})$ is the comoving sound horizon at $z_{*}$, and is given by Eq. (\ref{rs}).
From \cite{Planck2015,Neveu2016}, the $\chi^{2}$ is
\begin{equation}\label{X2CMB}
\chi_{\bf CMB}^{2}(\mathbf{X})=\left({\Delta x}_{i}\right)^{t}\left(C^{-1}_{\bf CMB}\right)_{ij}\left({\Delta x}_{j}\right),
\end{equation}
where ${\Delta x}_{i}=x^{\mathrm{th}}_{i}(\mathbf{X})-x^{\mathrm{obs}}_{i}$ is a column vector 
\begin{equation}
x^{\mathrm{th}}_{i}(\mathbf{X})-x^{\mathrm{obs}}_{i}=\left(\begin{array}{cc}
 l_{A}(z_{*})-301.7870\\
 R(z_{*})-1.7492\\
 \;\;z_{*}-1089.990\\
\end{array}\right),
\end{equation}
``t'' denotes its transpose and $(C^{-1}_{\bf CMB})_{ij}$ is the inverse covariance matrix \cite{Neveu2016}
given by
\begin{equation}\label{MatrixCMB}
C^{-1}_{\bf CMB}\equiv\left(
\begin{array}{ccc}
+162.48&-1529.4&+2.0688\\
-1529.4&+207232&-2866.8\\
+2.0688&-2866.8&+53.572\\
\end{array}\right).
\end{equation}
The errors for the CMB data are contained in $C^{-1}_{\bf CMB}$.
\subsubsection{Hubble data $\bar{H}(z)$}\label{OHD}
\begin{figure*}[!htb]
\centering
\resizebox{0.9\textwidth}{!}{
\begin{tabular}{*{2}{ c }}
\includegraphics{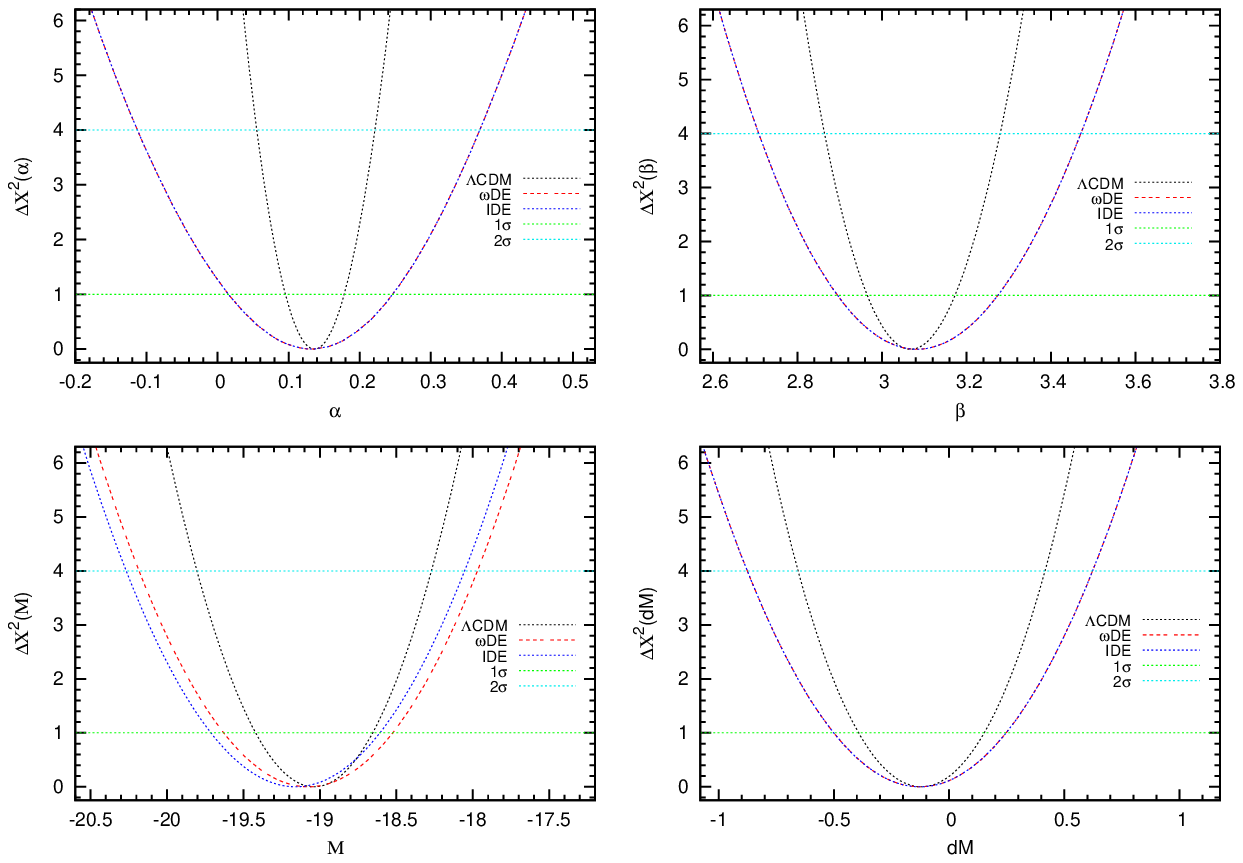}&\includegraphics{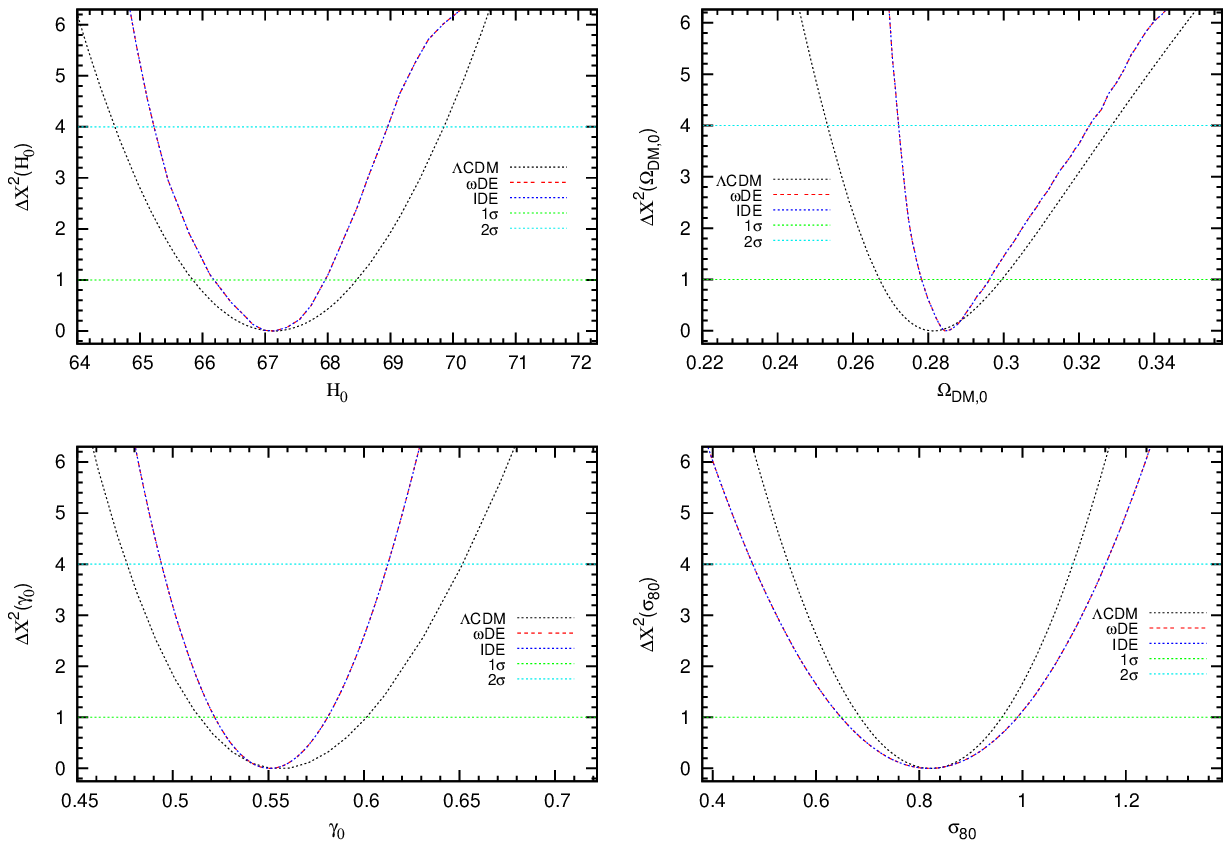}\\
\includegraphics{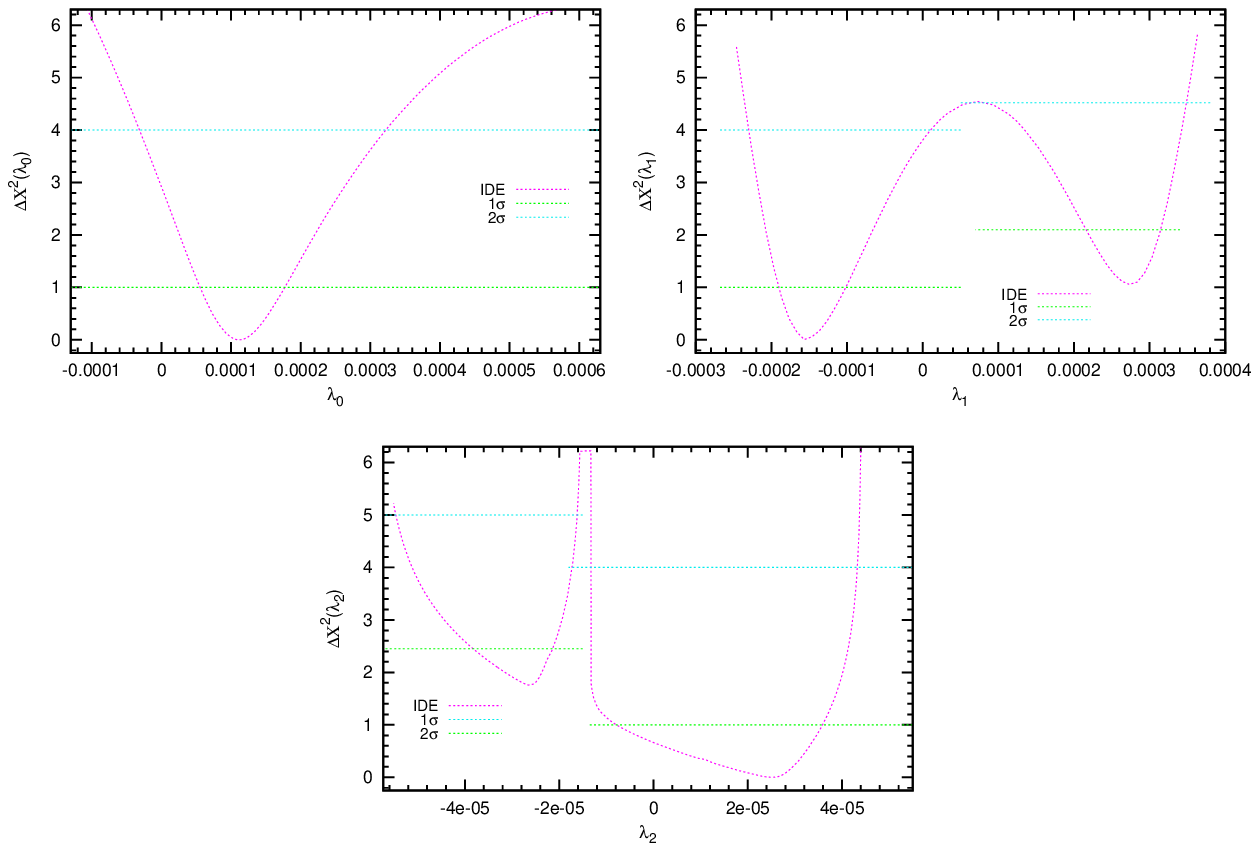}&\includegraphics{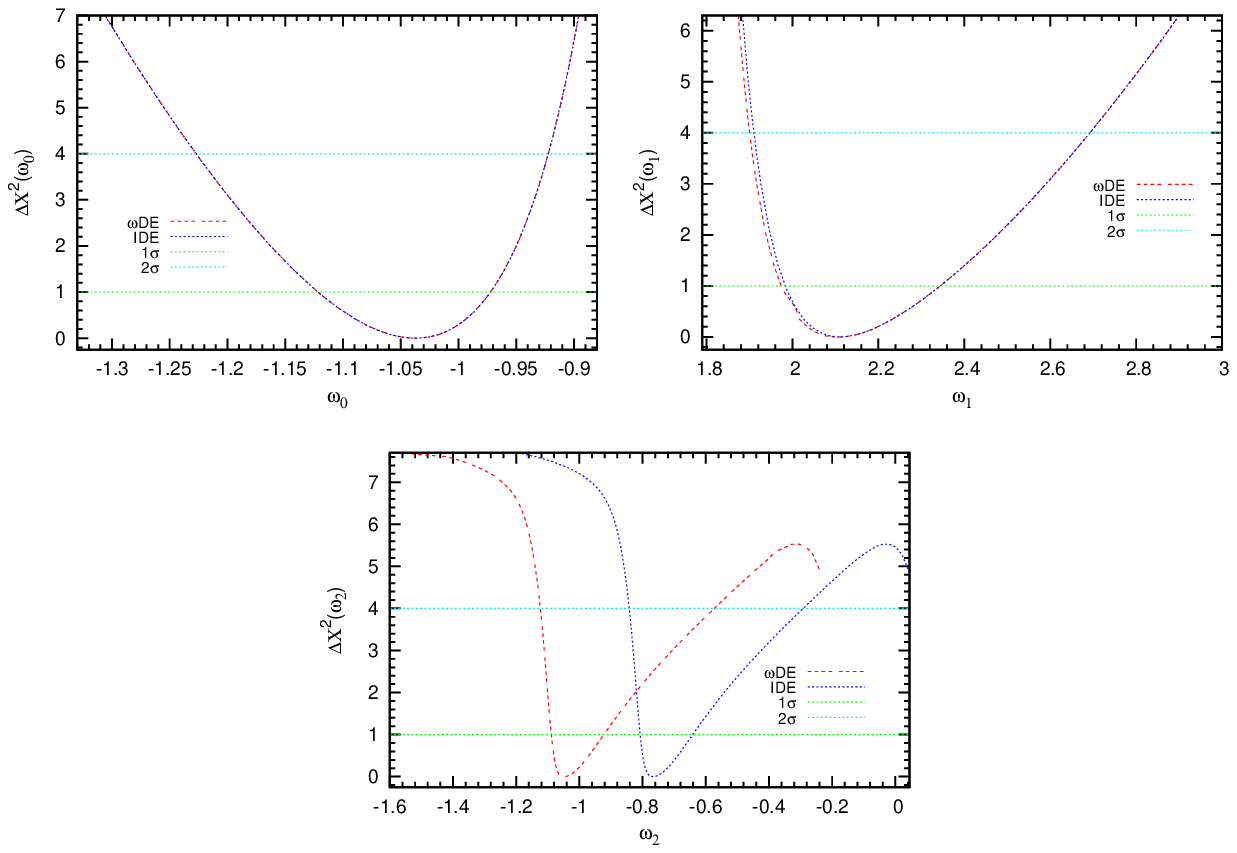} 
\end{tabular}} 
\caption{\label{Contours} Displays the one-dimension probability contours of the parameter space at $1\sigma$ and 
$2\sigma$ errors. Besides $\mathrm{\Delta{\chi^{2}}=\chi^{2}-{\chi^{2}_{min}}}$.}
\end{figure*}

\begin{figure*}[!htb]
 \centering 
 \resizebox{0.85\textwidth}{!}{
 \includegraphics{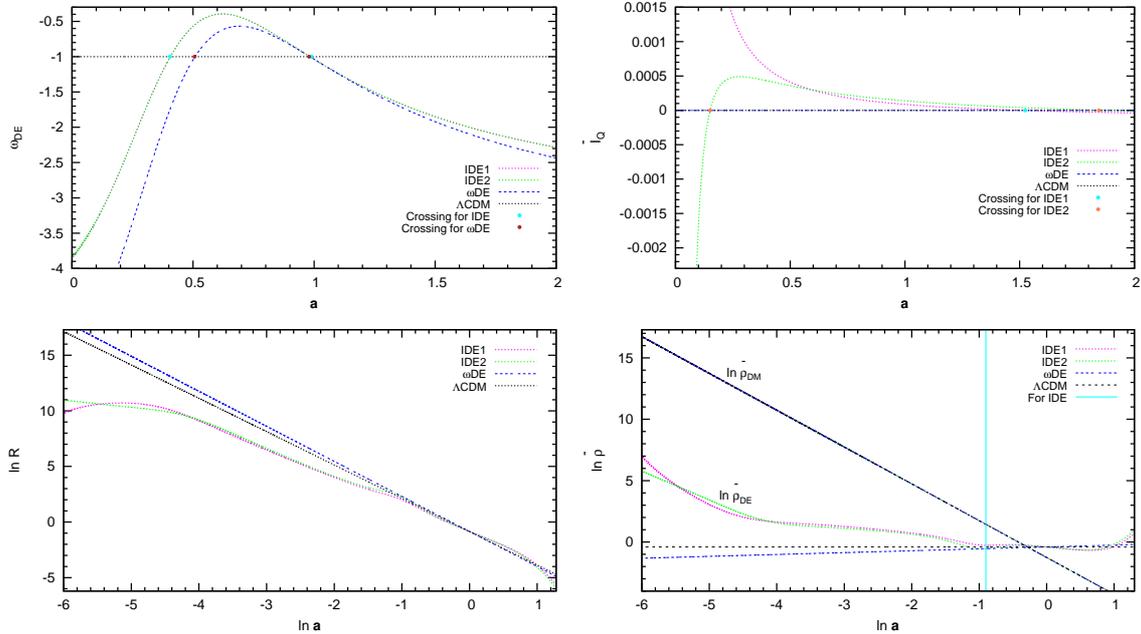}} 
 \caption{\label{Reconstructions} Shows the background evolution of $\mathrm{\omega_{DE}}$ and ${\bar{\rm I}}_{\rm Q}$, $\mathrm{R}$ and 
 $\mathrm{\bar{\rho}}$ along $\mathrm{a}$ for the coupled and uncoupled models. Here, we have fixed the best-fit values of Table \ref{Bestfits} 
 and have omitted the constraints at $1\sigma$ and $2\sigma$ to have a better visualization of the results.} 
\end{figure*}

This sample is composed by 38 independent measurements of the Hubble parameter at different redshifts \cite{Sharov2015} and were derived 
from differential age $dt$ for passively evolving galaxies with redshift $dz$ and from the two-points correlation function of Sloan Digital Sky Survey. 
This sample was taken from Table $\rm{III}$ in \cite{Sharov2015}. Then, the $\chi^2_{\mathrm{H}}$ function for this data set is \cite{Sharov2015}
\begin{equation}\label{X2OHD}
\chi^2_{\bf{{H}}}(\mathbf{X})\equiv\sum_{i=1}^{38}\frac{\left[{H}^{\rm {th}}(\mathbf{X},\mathrm{z_{i}})-{H}^{\mathrm{obs}}(\mathrm{z_{i}})\right]^2}{\sigma^2(\mathrm{z_{i}})},
\end{equation}
where ${H}^{{\rm th}}$ denotes the theoretical value of $\bar{H}$, ${H}^{\mathrm{obs}}$ re-presents its observed value and 
$\sigma(\mathrm{z_{i}})$ is the error.
\begin{table}[!htb]
\centering
\resizebox{0.48\textwidth}{!}{
\begin{tabular}{*{8}{|l}|}
\hline\noalign{\smallskip}
$\mathrm{z}$&$\mathrm{{f\sigma8}^{obs}}$&$\sigma$& Refs.&$\mathrm{z}$&$\mathrm{{f\sigma8}^{obs}}$&$\sigma$& Refs.\\ 
\noalign{\smallskip}\hline\noalign{\smallskip}
$0.020$&$0.360$&$\pm0.041$&\cite{Hudson2013}&$0.400$&$0.419$&$\pm0.041$&\cite{Tojeiro2012}  \\
$0.067$&$0.423$&$\pm0.055$&\cite{Beutler2012}&$0.410$& $0.450$&$\pm0.040$&\cite{Blake2011}\\
$0.100$&$0.370$&$\pm0.130$&\cite{Feix2015}&$0.500$&$0.427$&$\pm0.043$&\cite{Tojeiro2012}\\
$0.170$&$0.510$&$\pm0.060$&\cite{Percival2004}&$0.570$& $0.427$&$\pm0.066$&\cite{Reid2012}\\      
$0.220$&$0.420$&$\pm0.070$&\cite{Blake2011}&$0.600$&$0.430$&$\pm0.040$&\cite{Blake2011}\\
$0.250$&$0.351$&$\pm0.058$&\cite{Samushia2012}&$0.600$& $0.433$&$\pm0.067$&\cite{Tojeiro2012}\\ 
$0.300$&$0.407$&$\pm0.055$&\cite{Tojeiro2012}&$0.770$& $0.490$&$\pm0.180$&\cite{Song2009,Guzzo2008}\\
$0.350$&$0.440$&$\pm0.050$&\cite{Song2009,Tegmark2006}&$0.780$&$0.380$ &$\pm0.040$&\cite{Blake2011}\\
$0.370$&$0.460$&$\pm0.038$&\cite{Samushia2012}&$0.800$& $0.470$&$\pm0.080$&\cite{delaTorre2013}\\
\noalign{\smallskip}\hline
\end{tabular}}
\caption{Summary of RSD data set 
\cite{Hudson2013,Beutler2012,Feix2015,Percival2004,Song2009,Tegmark2006,Guzzo2008,Samushia2012,Blake2011,Tojeiro2012,Reid2012,delaTorre2013}.}\label{tableRSD}
\end{table}
\begingroup
\begin{table}[!htb]
\centering
\resizebox{0.45\textwidth}{!}{
\begin{tabular}{*{8}{|l}|}
\hline\noalign{\smallskip}
 $\mathrm{z}$ & $\mathrm{d_{z}^{obs}}$ & $\sigma_{z}$& Refs. & $\mathrm{z}$ & $\mathrm{d_{z}^{obs}}$ & $\sigma$& Refs.\\
\noalign{\smallskip}\hline\noalign{\smallskip}
$0.106$ & $0.3360$ &$\pm0.0150$ &\cite{Hinshaw2013,Beutler2011} &$0.350$ & $0.1161$ &$\pm0.0146$&\cite{Chuang2013a}\\
$0.150$ & $0.2232$ &$\pm0.0084$ &\cite{Ross2015} &$0.440$ & $0.0916$ &$\pm0.0071$&\cite{Blake2011}\\
$0.200$ & $0.1905$ &$\pm0.0061$ &\cite{Percival2010,Blake2011} &$0.570$ & $0.0739$ &$\pm0.0043$&\cite{Chuang2013b}\\
$0.275$ & $0.1390$ &$\pm0.0037$ &\cite{Percival2010} &$0.570$ & $0.0726$ &$\pm0.0014$&\cite{Anderson2014a}\\
$0.278$ & $0.1394$ &$\pm0.0049$ &\cite{Kazin2010} &$0.600$ & $0.0726$ &$\pm0.0034$&\cite{Blake2011}\\
$0.314$ & $0.1239$ &$\pm0.0033$ &\cite{Blake2011} &$0.730$ & $0.0592$ &$\pm0.0032$&\cite{Blake2011}\\
$0.320$ & $0.1181$ &$\pm0.0026$ &\cite{Anderson2014a}&$2.340$ & $0.0320$ &$\pm0.0021$&\cite{Debulac2015}\\
$0.350$ & $0.1097$ &$\pm0.0036$ &\cite{Percival2010,Blake2011} &$2.360$ & $0.0329$ &$\pm0.0017$&\cite{FontRibera2014}\\
$0.350$ & $0.1126$ &$\pm0.0022$ &\cite{Padmanabhan2012}        &$ $&$ $\\ 
\noalign{\smallskip}\hline
\end{tabular}}
\caption{Summary of BAO\,I data 
\cite{Blake2011,Hinshaw2013,Beutler2011,Ross2015,Percival2010,Kazin2010,Padmanabhan2012,Chuang2013a,Chuang2013b,Anderson2014a,Debulac2015,FontRibera2014}.}\label{tableBAOI}
\end{table}
\endgroup
\\
\begingroup
\begin{table}[!htb]
\centering
\resizebox{0.45\textwidth}{!}{
\begin{tabular}{*{8}{|l}|}
\hline\noalign{\smallskip}
 $\mathrm{z}$   & ${\bar{H}(z)}$ &  $1\sigma$& Refs. & $\mathrm{z}$   & ${\bar{H}(z)}$ &  $1\sigma$& Refs. \\
\noalign{\smallskip}\hline\noalign{\smallskip}
$0.070$&  $69.0$&  $\pm19.6$&\cite{Zhang2014}    & $0.570$&  $96.8$&  $\pm3.40$&\cite{Anderson2014a}\\
$0.090$&  $69.0$&  $\pm12.0$&\cite{Simon2005}    & $0.593$& $104.0$&  $\pm13.0$&\cite{Moresco2012}\\
$0.120$&  $68.6$&  $\pm26.2$&\cite{Zhang2014}    & $0.600$&  $87.9$&  $\pm6.1$ &\cite{Blake2012}\\
$0.170$&  $83.0$&  $\pm8.0$&\cite{Simon2005}     & $0.680$&  $92.0$&  $\pm8.0$ &\cite{Moresco2012}\\
$0.179$&  $75.0$&  $\pm4.0$&\cite{Moresco2012}   & $0.730$&  $97.3$&  $\pm7.0$ &\cite{Blake2012}\\
$0.199$&  $75.0$&  $\pm5.0$&\cite{Moresco2012}   & $0.781$& $105.0$&  $\pm12.0$&\cite{Moresco2012}\\
$0.200$&  $72.9$&  $\pm29.6$&\cite{Zhang2014}    & $0.875$& $125.0$&  $\pm17.0$&\cite{Moresco2012}\\
$0.240$&  $79.69$& $\pm2.99$&\cite{Gastanaga2009}& $0.880$&  $90.0$&  $\pm40.0$&\cite{Stern2010}\\
$0.270$&  $77.0$&  $\pm14.0$&\cite{Simon2005}    & $0.900$& $117.0$&  $\pm23.0$&\cite{Simon2005}\\
$0.280$&  $88.8$&  $\pm36.6$&\cite{Zhang2014}    & $1.037$& $154.0$&  $\pm20.0$&\cite{Gastanaga2009}\\
$0.300$&  $81.7$&  $\pm6.22$&\cite{Oka2014}      & $1.300$& $168.0$&  $\pm17.0$&\cite{Simon2005}\\
$0.340$&  $83.8$&  $\pm3.66$&\cite{Gastanaga2009}& $1.363$& $160.0$&  $\pm33.6$&\cite{Moresco2015}\\
$0.350$&  $82.7$&  $\pm9.1$& \cite{Chuang2013a}  & $1.430$& $177.0$&  $\pm18.0$&\cite{Simon2005}\\
$0.352$&  $83.0$&  $\pm14.0$&\cite{Moresco2012}  & $1.530$& $140.0$&  $\pm14.0$&\cite{Simon2005}\\
$0.400$&  $95.0$&  $\pm17.0$&\cite{Simon2005}    & $1.750$& $202.0$&  $\pm40.0$&\cite{Simon2005}\\
$0.430$&  $86.45$& $\pm3.97$&\cite{Gastanaga2009}& $1.965$& $186.5$&  $\pm50.4$&\cite{Moresco2015}\\
$0.440$&  $82.6$&  $\pm7.8$&\cite{Blake2012}     & $2.300$& $224.0$&  $\pm8.6$ &\cite{Busca2013}\\
$0.480$&  $97.0$&  $\pm62.0$&\cite{Stern2010}    & $2.340$& $222.0$&  $\pm8.5$ &\cite{Debulac2015}\\
$0.570$&  $87.6$&  $\pm7.80$&\cite{Chuang2013b}  & $2.360$& $226.0$&  $\pm9.3$ &\cite{FontRibera2014}\\
\noalign{\smallskip}\hline
\end{tabular}}
\caption{Shows the ${\bar{H}(z)}$ data 
\cite{Chuang2013a,Chuang2013b,Anderson2014a,Debulac2015,FontRibera2014,Zhang2014,Simon2005,Moresco2012,Gastanaga2009,Oka2014,Blake2012,Stern2010,Moresco2015,Busca2013}}\label{tableOHD} 
\end{table}
\section{Results}\label{Results}
\begingroup
\begin{table}[!htb]
\resizebox{0.35\textwidth}{!}{
\begin{tabular}{*{2}{|l}|}
\hline\noalign{\smallskip}
Parameters&Constant Priors\\[0.2mm]
\noalign{\smallskip}\hline\noalign{\smallskip}
$\mathrm{{\lambda}_{0}}$&$[-1.5\times10^{+2},+1.5\times10^{+2}]$\\[0.2mm]
$\mathrm{{\lambda}_{1}}$&$[-1.5\times10^{+2},+1.5\times10^{+2}]$\\[0.2mm]
$\mathrm{{\lambda}_{2}}$&$[-1.5\times10^{+1},+1.5\times10^{+1}]$\\[0.2mm]
$\mathrm{\omega_{0}}$&$[-2.0,-0.3]$\\[0.2mm]
$\mathrm{\omega_{1}}$&$[-1.0,+1.0]$\\[0.2mm]
$\mathrm{\omega_{2}}$&$[-2.0,+0.1]$\\[0.2mm]
$\mathrm{\Omega_{DM,0}}$&$[0,0.7]$\\[0.2mm]
$\mathrm{H_{0}}(\mathrm{kms^{-1}Mpc^{-1}})$&$[20,120]$\\[0.2mm]
$\mathrm{\alpha}$&$[-0.2,+0.5]$\\[0.2mm]
$\mathrm{\beta}$&$[+2.1,+3.8]$\\[0.2mm]
$\mathrm{M}$&$[-20,-17]$\\[0.2mm]
$\mathrm{dM}$&$[-1.0,+1.0]$\\[0.2mm]
$\mathrm{\gamma_{0}}$&$[+0.2,+1.2]$\\[0.2mm]
$\mathrm{\sigma_{80}}$&$[0,+1.65]$\\[0.2mm]
\noalign{\smallskip}\hline
\end{tabular}} 
\caption{Shows the priors on the parameter space.\hspace{2.7cm}}\label{Priors}
\end{table}
\endgroup
\begin{table*}[!hbtp]
\centering
\resizebox{0.85\textwidth}{!}{
\begin{tabular}{*{6}{|c}|}
\hline\noalign{\smallskip}
Parameters&$\Lambda$CDM&$\omega$DE&IDE1&IDE2\\
\noalign{\smallskip}\hline\noalign{\smallskip} 
$\mathrm{{\lambda}_{0}}\times10^{+4}$&$N/A$&$N/A$&${+1.120}^{+0.6525+2.0844}_{-0.6004-4.2761}$&${+1.120}^{+0.6525+2.0844}_{-0.6004-4.2761}$\\[0.4mm]
$\mathrm{{\lambda}_{1}}\times10^{+4}$&$N/A$&$N/A$&${+2.733}^{+0.4106+0.7486}_{-0.5670-1.8947}$&${+2.733}^{+0.4106+0.7486}_{-0.5670-1.8947}$\\[0.4mm]
$\mathrm{{\lambda}_{2}}\times10^{+5}$&$N/A$&$N/A$&${+2.539}^{+1.0254+1.7806}_{-10.2680-3.8349}$&${-2.649}^{+0.5207+1.0435}_{-1.1834-2.8293}$\\[0.6mm]
$\mathrm{\omega_{0}}$&$-1.0$&${-1.0364}^{+0.0644+0.1140}_{-0.0853-0.1908}$&${-1.0364}^{+0.0644+0.1140}_{-0.0853-0.1908}$&${-1.0364}^{+0.0644+0.1140}_{-0.0853-0.1908}$\\[0.4mm]
$\mathrm{\omega_{1}}$&$N/A$&${+2.0964}^{+0.2444+0.5970}_{-0.1228-0.1958}$&${+2.1064}^{+0.2363+0.5842}_{-0.1213-0.1964}$&${+2.1064}^{+0.2363+0.5842}_{-0.1213-0.1964}$\\[0.4mm]
$\mathrm{\omega_{2}}$&$N/A$&${-1.0510}^{+0.1326+0.4751}_{-0.0388-0.0712}$&${-0.7698}^{+0.1276+0.4797}_{-0.0364-0.0717}$&${-0.7698}^{+0.1276+0.4797}_{-0.0364-0.0717}$\\[0.4mm]
$\mathrm{\Omega_{DM,0}}$&${+0.2810}^{+0.0185+0.0476}_{-0.0138-0.0279}$&${+0.2844}^{+0.0121+0.0385}_{-0.0061-0.0124}$&${+0.2844}^{+0.0121+0.0385}_{-0.0061-0.0124}$&${+0.2844}^{+0.0121+0.0385}_{-0.0061-0.0124}$\\[0.4mm]
$\mathrm{\Omega_{b,0}}$&${+0.0493}^{+0.0018+0.0037}_{-0.0020-0.0040}$&${+0.0494}^{+0.0012+0.0025}_{-0.0014-0.0030}$&${+0.0494}^{+0.0012+0.0025}_{-0.0014-0.0030}$&${+0.0494}^{+0.0012+0.0025}_{-0.0014-0.0030}$\\[0.4mm]
$\mathrm{H_{0}}(\mathrm{kms^{-1}Mpc^{-1}})$&${+67.190}^{+1.3508+4.0403}_{-1.2203-3.3504}$&${+67.1490}^{+0.8216+1.8006}_{-0.9642-1.9324}$&${+67.1490}^{+0.8216+1.8006}_{-0.9642-1.9324}$&${+67.1490}^{+0.8216+1.8006}_{-0.9642-1.9324}$\\[0.4mm]
$\mathrm{\alpha}$&${+0.1360}^{+0.0419+0.0855}_{-0.0410-0.0814}$&${+0.1360}^{+0.1108+0.2341}_{-0.1198-0.2482}$&${+0.1360}^{+0.1108+0.2341}_{-0.1198-0.2482}$&${+0.1360}^{+0.1108+0.2341}_{-0.1198-0.2482}$\\[0.4mm]
$\mathrm{\beta}$&${+3.068}^{+0.1033+0.2129}_{-0.1026-0.2035}$&${+3.0780}^{+0.1968+0.3939}_{-0.1839-0.370}$&${+3.0780}^{+0.1968+0.3939}_{-0.1839-0.370}$&${+3.0780}^{+0.1968+0.3939}_{-0.1839-0.3700}$\\[0.4mm]
$\mathrm{M}$&${-19.0340}^{+0.3849+0.7605}_{-0.3907-0.7690}$&${-19.0810}^{+0.5674+1.1106}_{-0.5551-1.1043}$&${-19.1650}^{+0.5561+1.1116}_{-0.5522-1.0996}$&${-19.1650}^{+0.5561+1.1116}_{-0.5522-1.0996}$\\[0.4mm]
$\mathrm{dM}$&${-0.120}^{+0.0299+0.2983}_{-0.2718-0.5360}$&${-0.1250}^{+0.3326+0.6632}_{-0.3365-0.6633}$&${-0.1250}^{+0.3715+0.7488}_{-0.3832-0.7539}$&${-0.1250}^{+0.3715+0.7488}_{-0.3832-0.7539}$\\[0.4mm]
$\mathrm{\gamma_{0}}$&${+0.5511}^{+0.0506+0.1010}_{-0.0375-0.0753}$&${+0.5511}^{+0.0302+0.0615}_{-0.0291-0.0571}$&${+0.5511}^{+0.0302+0.0615}_{-0.0291-0.0571}$&${+0.5511}^{+0.0302+0.0615}_{-0.0291-0.0571}$\\[0.4mm]
$\mathrm{\sigma_{80}}$&${+0.8180}^{+0.1400+0.2794}_{-0.1340-0.2718}$&${+0.8190}^{+0.1706+0.3425}_{-0.1690-0.3417}$&${+0.8190}^{+0.1706+0.3425}_{-0.1690-0.3417}$&${+0.8190}^{+0.1706+0.3425}_{-0.1690-0.3417}$\\[0.4mm]
$\mathrm{\tau}$&${+0.0660}^{+0.0012+0.0033}_{-0.0013-0.0026}$&${+0.0661}^{+0.0008+0.0017}_{-0.0010-0.0020}$&${+0.0661}^{+0.0008+0.0017}_{-0.0010-0.0020}$&${+0.0661}^{+0.0008+0.0017}_{-0.0010-0.0020}$\\[0.4mm]
$\mathrm{\Theta_{s}}$&${+1.0479}^{+0.0164+0.0367}_{-0.01525-0.0310}$&${+1.0493}^{+0.0106+0.0274}_{-0.0092-0.0188}$&${+1.0493}^{+0.0106+0.0274}_{-0.0092-0.0188}$&${+1.0493}^{+0.0106+0.0274}_{-0.0092-0.0188}$\\[0.4mm]
$\mathrm{\ln[10^{10}A_{s}]}$&${+4.5307}^{+0.0324+0.0809}_{-0.0245-0.0508}$&${+4.4314}^{+0.0286+0.0858}_{-0.0152-0.0316}$&${+4.4314}^{+0.0286+0.0858}_{-0.0152-0.0316}$&${+4.4314}^{+0.0286+0.0858}_{-0.0152-0.0316}$\\[0.4mm]
\hline\hline${\chi}^{2}_{min}$&$737.8581$&$726.0120$&$712.3048$&$723.7758$\\[0.4mm]
\noalign{\smallskip}\hline
\end{tabular}}
\caption{Shows the best-fit values of the cosmological parameters for the studied models with $1\sigma$ and $2\sigma$ errors.}\label{Bestfits}
\end{table*}
\endgroup
In this work, we have ran eight chains for each of our models on the computer, and the obtained outcomes of the main and derived parameters are 
presented in Table \ref{Bestfits}, in where the best estimated parameters with their $1\sigma$ and $2\sigma$ errors are shown. Moreover, the minimum 
${\chi}^{2}_{min}$ is $712.3048$ for the IDE model, which is smaller in comparison with those obtained in the non-interacting models 
and the one-dimension probability contours at $1\sigma$ and $2\sigma$ on single parameters are plotted in Fig. \ref{Contours}.\\
Likewise, from Table \ref{Bestfits} and Fig. \ref{Contours}, we notice that the inclusion of CMB and RSD data allow to break the degeneracy among 
the different parameters of our models, obtaining constraints more stringent on them. When ${\bar{\rm I}}_{\rm Q}=0$, one finds that the 
$\omega$DE model is very close to the IDE model.\\
Due to the two minimums obtained in the IDE model (see Table \ref{Bestfits}), we consider now two different cases to recons-truct ${\rm I}_{\rm Q}$: 
the case 1 is so-called IDE1 with ${\lambda}_{2}>0$; by contrast, the case 2 is so-called IDE2 with ${\lambda}_{2}<0$.\\
From the left above panel of Fig. \ref{Reconstructions}, one can see that the universe evolves from the phantom regime $\mathrm{\omega_{DE}} < -1$ 
to the quintessence regime $\mathrm{\omega_{DE}} > -1$, and then it becomes phantom again; and in particular, crosses the phantom divide line 
$\mathrm{\omega_{DE}}=-1$ \cite{Nesseris2007}. The IDE model has two crossing points in $\mathrm{\bf a}=0.4043$ and $\mathrm{\bf a}=0.9894$, 
respectively. Such a crossing feature $\mathrm{\omega_{DE}}=-1$ is favored by the data about at $1\sigma$ error. Then, our fitting results show that 
the evolution of $\mathrm{\omega_{DE}}$ in the $\omega$DE and IDE models are very close to each other, in particular, they are close to $-1$ today.\\ 
Now, in the right above panel of Fig. \ref{Reconstructions}, we have considered that at early times when DM dominates the universe
${\rm I}_{+}$ denotes an energy transfer from DE to DM and ${\rm I}_{-}$ denotes an energy transfer from DM to DE. Here, we have found a change from 
${\rm I}_{+}$ to ${\rm I}_{-}$ and vice versa. This change of sign is linked to ${\bar{\rm I}}_{\rm Q}=0$ and is also favored by the data at 
$1\sigma$ error. The IDE model shows three crossing points in $\mathrm{\bf a}=1.5238$ (IDE1), $\mathrm{\bf a}=0.1512$ (IDE2) and 
$\mathrm{\bf a}=1.8462$ (IDE2), respectively. The fitting results indicate that ${\rm I}_{\rm Q}$ is stronger at early times and weaker at later times, 
namely, ${\rm I}_{\rm Q}$ remains small today, being ${\rm I}_{\rm Q,0}={0.8661\times10^{-4}}^{+0.55\times10^{-4}}_{-0.4977\times10^{-4}}$ for the 
case IDE1 and ${\rm I}_{\rm Q,0}={1.3849\times10^{-4}}^{+0.6044\times10^{-4}}_{-0.4820\times10^{-4}}$ for the case IDE2, respectively. These results 
are consistent at $1\sigma$ error with those reported in \cite{Cueva-Nucamendi2012,Cai2010,Li2011}. However, our outcomes are smaller. This small 
discrepancy is due to the ansatzes chosen for ${\rm I}_{\rm Q}$ and the data used.\\
Also, in the left below panel of Fig. \ref{Reconstructions}, we note that $\mathrm{R}$ is always positive when both $\bar{{\rm I}}_{\rm Q}$ and 
$\mathrm{\omega_{DE}}$ are time-varying, and remains finite when $\mathrm{\bf a}\rightarrow \infty$. As is apparent, $\bar{Q}$ seems to alleviate 
the coincidence problem for ln $\mathrm{\bf a}\leq 0$. Likewise, from the right below panel of this figure, we note that the vertical line 
indicates the moment when $|3\bar{\mathcal{H}}(1+\mathrm{\omega_{DE}})\mathrm{{\bar\rho}_{DE}}|$ and 
$|\mathrm{{\bar\rho}_{DM}}{\bar{\rm I}}_{\rm Q}|$ are equal, see Eq. (\ref{EDE}). Here, to the left of this line, $\bar{Q}$ affects the background 
evolution of $\mathrm{\bar{\rho}_{DE}}$. By contrast, the situation is opposite to the right of this line.
This panel also shows that the background evolution of $\mathrm{\ln R}$, exhibits a scaling behavior at early 
times (keeping constant) but not at the present day. These results signifi-cantly alleviate the coincidence problem, but they do not 
solve it in full. From the right below panel of the Fig. \ref{Reconstructions}, we see that at $\mathrm{\ln{\bf a}}<0$ the coupling affects 
violently the background evolution of $\mathrm{{\bar{\rho}}_{DE}}$ in the IDE model. By contrast, the situation is opposite at 
$\mathrm{\ln \bf{a}}>0$. Furthermore, the graphs for $\mathrm{{\bar{\rho}}_{DM}}$ are essentially overlapped during their evolution.
 \begin{figure*}[!htb]
 \centering 
 \resizebox{0.85\textwidth}{!}{
 \includegraphics{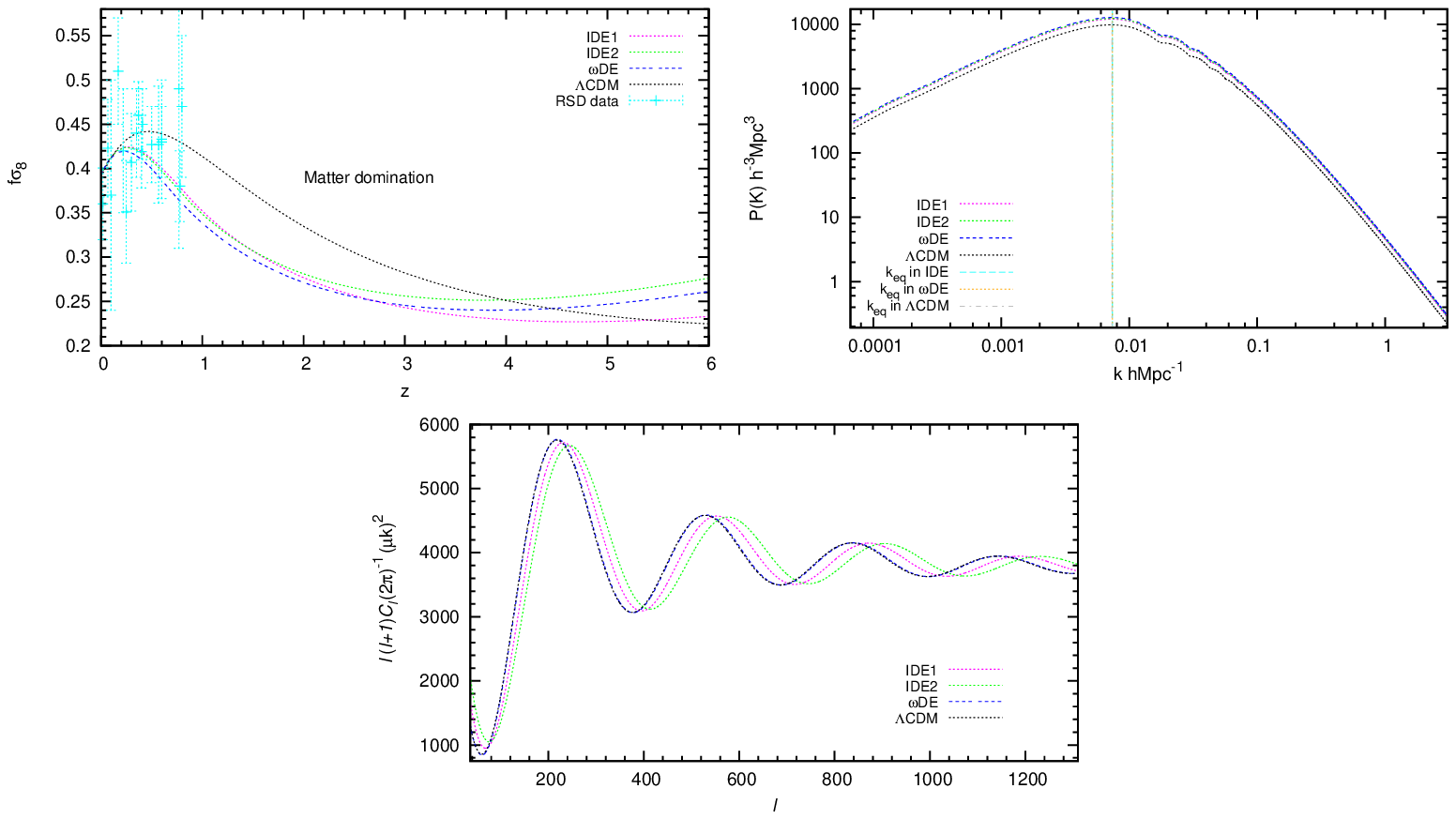}} 
  \caption{\label{Effects1} Shows the combined impact of $\bar{Q}$ and $\mathrm{\omega_{DE}}$ on the evolution of the structure growth of the matter 
  $\mathrm{f\sigma_{8}}$, on the linear matter power spectrum $\mathrm{P(k)}$, and on the CMB temperature power spectrum 
  $\it{l(l+1)C_{l}}\mathrm{(2\pi)^{-1}}\mathrm{(\mu k)^{2}}$, respectively.}
 \end{figure*}
 \\
 The left above panel of Fig. \ref{Effects1}, shows the evolution of the structure growth of the matter, $\mathrm{f\sigma_{8}}$, along 
 $\mathrm{z}$ for the different cosmologies. These curves are comparable with each other at $\mathrm{z<0.4}$ but they deviate one after another at 
 $\mathrm{z>0.4}$. It implies that they are sensitive to the background cosmology. Within the matter era the amplitude of $\mathrm{f\sigma_{8}}$ in the 
 IDE model is enhanced relative to the $\omega$DE model at $\mathrm{z}<3$, but both are smaller than that found in the $\Lambda$CDM. At $\mathrm{z}>5$, 
 $\bar{Q}$ and $\mathrm{\omega_{DE}}$ would brough about a large structure formation in the IDE and $\omega$DE scenarios, respectively. Due to the fact that 
 the amount of DM is bigger than the amount of DE at earlier times; therefore, it produces an enhancement on the amplitudes of $\mathrm{f\sigma_{8}}$ 
 in the IDE model respect to that found in the $\Lambda$CDM, respectively. Our fitting results are consistent at $1\sigma$ error with those reported by 
 \cite{Mehrabi2015,Clemson2012,Alcaniz2013,Yang2014,Mota2017}.
 The right above panel of this Figure depicts the evolution of the total matter power for different scenarios at $\mathrm{z=1.6}$. Notice that $\mathrm{P(k)}$ in the 
 IDE model is enhanced with respect to that in the $\Lambda$CDM  but it is suppressed in relation to that found in the $\omega$DE scenario. That could be 
 a consequence of the amount of concentrated matter at early times and also the presence of $\bar{Q}$. Moreover, the vertical line indicates the turnover 
 position, $\mathrm{k_{eq}}$, is very close in our models. Also, we notice a series of wiggles on the $\mathrm{P(k)}$ due 
 to the coupling between the photons and baryons before recombination; namely, the presence of baryons have left their effect there.  these 
 arguments, the IDE and $\omega$DE can be distinguished from the curves of $\mathrm{f\sigma_{8}}$, and the structure growth data could break the 
 possible degeneracy between these two models and provides a signature to discriminate them. These outcomes are in correspondence with those found 
 by \cite{Clemson2012,Yang2014,Mota2017} at $1\sigma$ error. 
 Likewise, the below panel of this Figure displays the effects of $\bar{Q}$ and $\mathrm{\omega_{DE}}$ on the amplitude of the CMB temperature power 
 spectrum at low multipoles $\textit{l}<100$, in where the amplitude of the integrated Sachs-Wolfe effect is deviated in the IDE model respect to that 
 found in the other scenarios. Instead, at high multipoles ${\textit{l}>100}$, $\bar{Q}$ and $\mathrm{\omega_{DE}}$ increase the concentration of DM 
 early times, affecting the sound horizon at the end, which shiftes to right the values of the acoustic peaks located at 
 $\textit{l}=n_{p}\pi D_{A}(z_{*})/r_{s}(z_{*}),\, n_{p}=1,2,3..$, and reduced the amplitudes of the first peaks when the studied models are compared. 
 These features can be understood by considering the extra-terms proportionals to $\mathrm{{\lambda}_{0}}$, $\mathrm{{\lambda}_{1}}$ and 
 $\mathrm{{\lambda}_{2}}$ in the  $\mathrm{\Omega_{DM}}$, which increases and, in consequence, amplifies the amount of DM at early times. That is in accordance with the result found in the 
 previous panel of this Figure and with those found in \cite{Clemson2012,Yang2014,Mota2017} at $1\sigma$ error.
 \section{Conclusions}\label{SectionConclusions}             
In this work, we examined an interacting DE model with an interaction $\bar{Q}$ proportional to the DM energy density, to the Hubble parameter $\bar{H}$, 
and to a time-varying function, ${\bar{\rm I}}_{\rm Q}$ expanded in terms of the Chebyshev polynomials $T_{n}$, defined in the interval $[-1,1]$. 
Besides, we also consider a time-varying EoS parameter, $\mathrm{\omega_{DE}}$, expressed in function of those polynomials. These ansatzes have 
been proposed so that their background evolution are free of divergences at the present time and also at the future time, respectively.
In a Newtonian gauge and on sub-horizon scales, a set of perturbed equations is obtained when the momentum transfer potential becomes null in the 
DM rest-frame. This leads to different cases in the IDE model. Based on a combined analysis of geometric and dynamical probes which include 
JLA + RSD + BAO + CMB + H data and using the MCMC, we found the best-fit parameters that constrain the background evolution of our models.\\ 
We have also considered the perturbed equations for the DM and baryons in the rest-frame of DM. Besides, we have built the theoretical and numerical 
structures, and in particular, we used the c++ language to show the combined impact of both $\bar{Q}$ and $\mathrm{\omega_{DE}}$ on the evolution of 
$\mathrm{R}$, $\mathrm{{\bar{\rho}}_{DM}}$, $\mathrm{{\bar{\rho}}_{DE}}$, $\mathrm{f\sigma_{8}}$, $\mathrm{P(k)}$ and $\sl{l(l+1)C_{l}}$, 
respectively.\\
Likewise from Table \ref{Bestfits} and Fig. \ref{Reconstructions}, our fitting results show that ${\bar{\rm I}}_{\rm Q}$ can cross twice the line $\bar{Q}=0$ during its background evolution. Similarly, 
$\mathrm{\omega_{DE}}$ crosses the line $\mathrm{\omega_{DE}}=-1$ twice as well. These crossing features are favored by the data at $1\sigma$ error. 
Furthermore, we also notice that $\mathrm{R}$ is always positive and remains finite in all our models when $\mathrm{\bf a}\rightarrow \infty$. Moreover, 
in the IDE model, $\mathrm{R}$ exhibits a sca-ling behaviour at early times (keeping constant). Then, $\bar{Q}$ seems to alleviate the coincidence 
problem for ln $\mathrm{\bf a}\leq 0$ but it does not solve that problem in full.\\
On the other hand, from Fig. \ref{Effects1}, we found that the evolution curve of $\mathrm{f\sigma_{8}}$ in the IDE model deviates significantly from that obtained in the 
$\Lambda$CDM and $\omega$DE models. It meant that, the structure formation data could break the possible degenaracy between the IDE and $\omega$DE models. 
In these last two models, several best-fit parameters are very close with each other, therefore, one could then conclude that this detected deviation 
is brough about mainly by $\mathrm{\omega_{DE}}$, namely, $\mathrm{f\sigma_{8}}$ is sensitive mainly to the evolution of $\mathrm{\omega_{DE}}$ and 
depends on its parametrisation. Moreover, the geometric probes favor the existence of an interaction between the dark sectors but the dynamical test 
constrains its intensity. These effects can be understood by conside-ring the extra-terms proportional to ${\bar{\rm I}}_{\rm Q}$ in the DM ener-gy density, 
(see Eq. (\ref{hubble})), which increases and, in consequence, amplifies the amount of DM at earlier times. As a result, the growth of structure is 
significantly affec-ted by $\bar{Q}$ and $\mathrm{\omega_{DE}}$, which induce that the amplitude of $\mathrm{P(k)}$ becomes higher in the IDE model than 
that found in the $\Lambda$CDM model but it is lesser than that found in the $\omega$DE. Moreover, the position of the turnover point in all our models 
is very close at smaller scales. Likewise, we also notice a series of wiggles on the curve of $\mathrm{P(k)}$ due to the coupling between the photons and 
baryons before of the recombination; namely, the presence of baryons have left their effect in this plot (see Fig. \ref{Effects1}). Finally, we also find 
that the amplitude of the CMB temperature power spectrum is also sensitive to $\bar{Q}$ at low and high multipoles. In the IDE model, $\bar{Q}$ produces 
a shift of the acoustic peaks to the right and their amplitudes are reduced at high multipoles with respect to the uncoupled models. Besides, at low 
multipoles the amplitude of the integrated Sachs-Wolfe effect is also affected by $\bar{Q}$.\\
The results for the other case when the momentum-transfer potential vanishing in the DE rest-frame will be presented in a future work.
\\\\
\begin{scriptsize}
\textbf{\textsf {Acknowledgments}}
\\
\textsf{The author is indebted to the Instituto de F\'{\i}sica y Matem\'aticas (UMSNH) for its hospitality and support.}
\end{scriptsize}

\end{document}